\documentclass{aa}  

\usepackage{graphicx}
\usepackage[varg]{txfonts}

\usepackage{color}
\usepackage[dvipsnames]{xcolor}

\newcommand{\halpha}{H$\alpha$}
\newcommand{\hbeta}{H$\beta$}
\newcommand{\lya}{Ly$\alpha$}
\newcommand{\HII}{\ion{H}{ii}}
\newcommand{\msun}{M$_{\sun}$}
\newcommand{\kms}{km~s$^{-1}$}

\newcommand{\sii}{[\ion{S}{II}]}
\newcommand{\nii}{[\ion{N}{II}]}
\newcommand{\oi}{[\ion{O}{I}]}
\newcommand{\oiii}{[\ion{O}{III}]}
\newcommand{\heii}{\ion{He}{ii}}
\begin{document}

   \title{
   Super star cluster feedback driving  ionization, shocks and outflows in the halo of the nearby starburst ESO 338-IG04\thanks{Based on observations made with ESO Telescopes at the La Silla Paranal Observatory under programme  60.A-9314 (Science verification) and on public data released from the MUSE commissioning observations at the VLT Yepun (UT4) telescope under Programme ID 60.A-9100(C).}}

\titlerunning{Super star cluster feedback in ESO338-IG04}

   \author{A. Bik
        \inst{1}
        \and
          G. \"Ostlin
          \inst{1}
        \and
	V. Menacho
	\inst{1}
	\and
	A. Adamo
	\inst{1}
	\and
	M. Hayes
	\inst{1}
	\and
	E.C. Herenz
	\inst{1}
	\and
    J. Melinder
	\inst{1}
	  }
   \institute{Department of Astronomy, Stockholm University, Oscar Klein Centre, AlbaNova University Centre, 106 91 Stockholm, Sweden\\
              \email{arjan.bik@astro.su.se}             
}

   \date{Received ; accepted }

  \abstract
   {Stellar feedback strongly affects  the interstellar medium (ISM) of galaxies.  Stellar feedback in the first galaxies likely plays a major role in enabling the escape of  LyC photons, which contribute to the re-ionization of the Universe. Nearby starburst galaxies serve as local analogues allowing for a spatially resolved assessment of the feedback processes in these galaxies.}
   {We characterize the feedback effects from the  star clusters in the local high-redshift analogue ESO 338-IG04 on the ISM and compare the results with the properties of the most massive clusters.}
   {We use high quality  VLT/MUSE optical integral field data to derive the physical properties of the ISM such as ionization, density, shocks, and perform new fitting of the spectral energy distributions of the brightest clusters in ESO 338-IG04 from HST imaging.}
   {ESO 338-IG04 has a large ionized halo which we detect to a distance of 9 kpc. We identify 4 Wolf-Rayet (WR) clusters based on the blue and red WR bump. We follow previously identified ionization cones and find that the ionization of the halo increases with distance. Analysis of the galaxy kinematics shows two complex outflows driven by the numerous young clusters in the galaxy.   We find a ring of shocked emission traced by an enhanced \oi/\halpha\ ratio surrounding the starburst and at the end of the outflow. Finally we detect nitrogen enriched gas associated with the outflow, likely caused by the WR stars in the massive star clusters.}
   {Photo-ionization dominates the central starburst and sets the ionization structure of the entire halo, resulting in a density bounded halo, facilitating the escape of LyC photons.  Outside the central starburst, shocks triggered by an expanding super bubble become important. The shocks at the end of the outflow suggest interaction between the hot outflowing material and the more quiescent halo gas.}

   \keywords{galaxies: starburst - galaxies: halos -  galaxies: kinematics and dynamics - galaxies: individual: ESO 338-IG04 - techniques: imaging spectroscopy}

   \maketitle
%

\section{Introduction}

Young massive stars have a strong impact on their surroundings. They destroy the molecular clouds they are born in and ionize the surrounding interstellar medium  via  copious amount of Lyman continuum (LyC) photons. Additionally, stellar winds and, at a later stage supernova explosions inject large amounts of mechanical energy in  the ISM, driving shocks and large scale outflows.  Stellar feedback has strong effects on the evolution of the galaxies where the massive stars are located in. Feedback can trigger new sites of star formation \citep[e.g.][]{Adamo12}, or even suppress star formation by emptying large cavities around massive star clusters  \citep[e.g.][]{Ostlin07,Pasquali11}.

At low-metallicity especially, the effect of the stellar feedback on the ISM of the galaxy is not well understood. A detailed understanding of stellar feedback in these environments and how it facilitates the escape of LyC radiation is particularly relevant as the leakage of the LyC from starburst galaxies is one of the candidates for the re-ionization of the universe \citep[e.g.][]{Robertson15,Bouwens15,Trebitsch17}.  
However, at high redshift these galaxies are hard to detect, let alone study in detail. Therefore studies of local galaxies with similar properties as the high-redshift star forming galaxies provide vital information in understanding how the feedback acts on the ISM and facilitates possible LyC leakage. 

Blue compact (dwarf) galaxies (BCGs) are typically considered to be local analogues of high redshift star forming galaxies. They are compact, characterized by a highly elevated star formation rate and low metallicity  \citep{Searle72,Izotov99}. These galaxies typically have an ongoing starburst super imposed on a much older stellar population  \citep[][]{Loose86,Kunth88,Papaderos96b,Papaderos96a,Kunth00,Bergvall02,GildePaz03}. Their optical spectra are dominated by emission lines allowing a detailed characterization of the ISM properties and kinematics. Kinematical studies of BCGs  reveal irregular velocity fields suggestive of their starburst being triggered by mergers involving dwarf galaxies or in-falling massive gas clouds \citep{Ostlin01}. BCGs are hosts of rich population of young stellar clusters \citep[e.g.][]{Meurer95,Thuan97,Ostlin99,Ostlin03,Adamo10,Adamo10SBS,Adamo11,Calzetti15}, containing large amounts of massive stars responsible for stellar feedback.

During the evolution of a star cluster different processes are responsible for the energy output that modify the ISM. According to stellar population models star clusters younger than $\sim$4 Myrs (for metallicity Z=0.004) emit the most amount of  Hydrogen and Helium ionizing photons as even the most massive stars have not yet exploded  as supernova \citep{Leitherer99}. In this phase ionization and radiation pressure are important mechanisms of feedback \citep{Krumholz14PPVI}.  After 3 Myrs massive stars evolve into Wolf-Rayet (WR) stars. Their surface temperatures can reach values up to 100,000 K \citep{Smith02,Crowther07} and they emit large amounts of very hard photons, giving rise to the ionization of He$^{+}$.  At low metallicity (12+log O/H < 8), however, evidence is found for a decreased importance of WR stars and that O stars could be the main source of He$^{+}$ ionizing photons\citep{Kunth86,Guseva00,Brinchmann08}. 

The strong stellar winds of massive stars are responsible for the output of mechanical energy in the ISM during the first 4 Myr of the cluster lifetime.  Only after 4  Myrs, when the first supernovae explode, mechanical energy is mainly produced by the supernovae. Until clusters reach an age of 40 Myrs, their mechanical energy output (by stellar wind and later by supernovae) remains roughly constant \citep{Leitherer99}. The mechanical energy of the stellar winds and supernovae heats up the surrounding ISM and creates a large, so called super-bubble, filled with hot, X-ray emitting gas \citep{Weaver77}. In the theoretical picture of the formation of galactic winds \citep{Chevalier85,Strickland09,Heckman17}, super-bubbles are the first stage in the creation of a galactic scale outflow. In the super bubble the reservoir of hot gas thermalised by the supernova explosions is built up. While expanding, the edge of the bubble becomes gravitationally unstable and break. This allows the hot gas to escape and form an outflow (the so-called 'blow-out" phase).  In a inhomogeneous medium, the outflowing material follows the path of least resistance and escapes via the low-density channels already present in the gas, created by e.g. previous generations of clusters \citep{Cooper08}.  The \halpha\ emission is originating from the warm gas entrained by the outflow and the inner walls of the outflow cones \citep[e.g.][]{Westmoquette09}, as well as from radiatively cooling hot gas \citep{Thompson16}.

In this paper we focus on one of the closest luminous BCG  and high redshift analogue ESO 338-IG04 (hereafter shortened to ESO 338, $\alpha$=19$^{\rm{h}}$27$^{\rm{m}}$58.2$^{\rm{s}}$, $\delta$=-41\degr 34\arcmin 32\arcsec), also known as Tololo 1924-416, at a distance of 37.5 Mpc\footnote{Assuming $v_{helio}$ = 2831 \kms and $H_0$ = 75 \kms} \citep{Ostlin98} and a metallicity of 12\% solar \citep{Bergvall85}.
The galaxy is a compact galaxy, currently undergoing a vigorous starburst which started 40 Myrs ago and created a large young stellar cluster (YSC) population \citep{Ostlin98,Ostlin03,Adamo11}.  The  most massive young stellar cluster  in this galaxy (cluster 23) has a dynamical mass of $\sim$10$^7$ \msun\ and is the most luminous and massive young (age $<$ 10 Myr) cluster detected in nearby galaxies. This cluster has evacuated a large bubble in the ISM \citep{Ostlin07}. ESO 338 has a total stellar mass of $4\times 10^9$ \msun\ and a star formation rate of 3.2 \msun\ yr$^{-1}$ \citep{Ostlin01}. The galaxy is surrounded by a large halo of ionized  \citep{Ostlin99,Bik15} and neutral gas \citep{Cannon04}.  ESO 338 has a companion galaxy, ESO 338-IG04B,  connected via a tidal tail observed in \ion{H}{I} \citep{Cannon04}. \citet{Cannon04} suggests that a strong gravitational interaction between the two galaxy has triggered a starburst in ESO 338-IG04. However, the projected distance is rather large ($\sim$72 kpc), suggesting that the current starburst may not have been triggered by this interaction.
Summarizing, the  relative proximity and extreme properties make ESO 338 a good environment to study the effect of cluster feedback at low metallicity. 

We present deep  optical integral field spectroscopy obtained with the Multi-Unit Spectroscopic Explorer (MUSE)  of this galaxy where we observe the ISM and the extended halo of the galaxy. This data will allow us to obtain spatially resolved information on physical properties of the ISM such as density, extinction, degree of ionization, abundance and kinematics. In \citet{Bik15} the first results of the MUSE observations on ESO 338 were presented, where we identified two ionization cones as well as two galactic scale  outflows. We found that the \lya\ emission preferentially escapes in the direction of the galactic scale outflows. This paper presents even deeper data by combining two different MUSE datasets on ESO 338.  This unique dataset from  MUSE enables us for the first time  to make a direct comparison between the properties of the ISM and the  young stellar cluster population in ESO 338 as derived from HST.

In this paper we combine the cluster analysis and the analysis of the ISM of ESO 338 in order to analyse the effect clusters have on the ISM of ESO 338 in detail. In Sect. \ref{sec:observations}, the data are presented and the data reduction is described. Sect. \ref{sec:clusters} summarises the properties of the stellar clusters and especially those of the clusters containing WR stars. In Sect. \ref{sec:gasprop} we derive the physical conditions of the ionized gas based on the emission line maps. In Sect. \ref{sec:halo} we analyze the halo in further detail, by deriving its mass, apply a two dimensional  analysis of emission line ratios using the so called BPT diagram \citep{Baldwin81}, as well as analyse the galactic scale outflow in detail. In Sect.  \ref{sec:discussion}  we discuss the properties of the halo in the context of other galaxies and speculate on the possibility of ESO 338 being a LyC leaking galaxies. The paper ends with a summary and conclusions (Sect. \ref{sec:conclusions}).

  \begin{figure}[!t]
   \centering
   \includegraphics[width=\hsize]{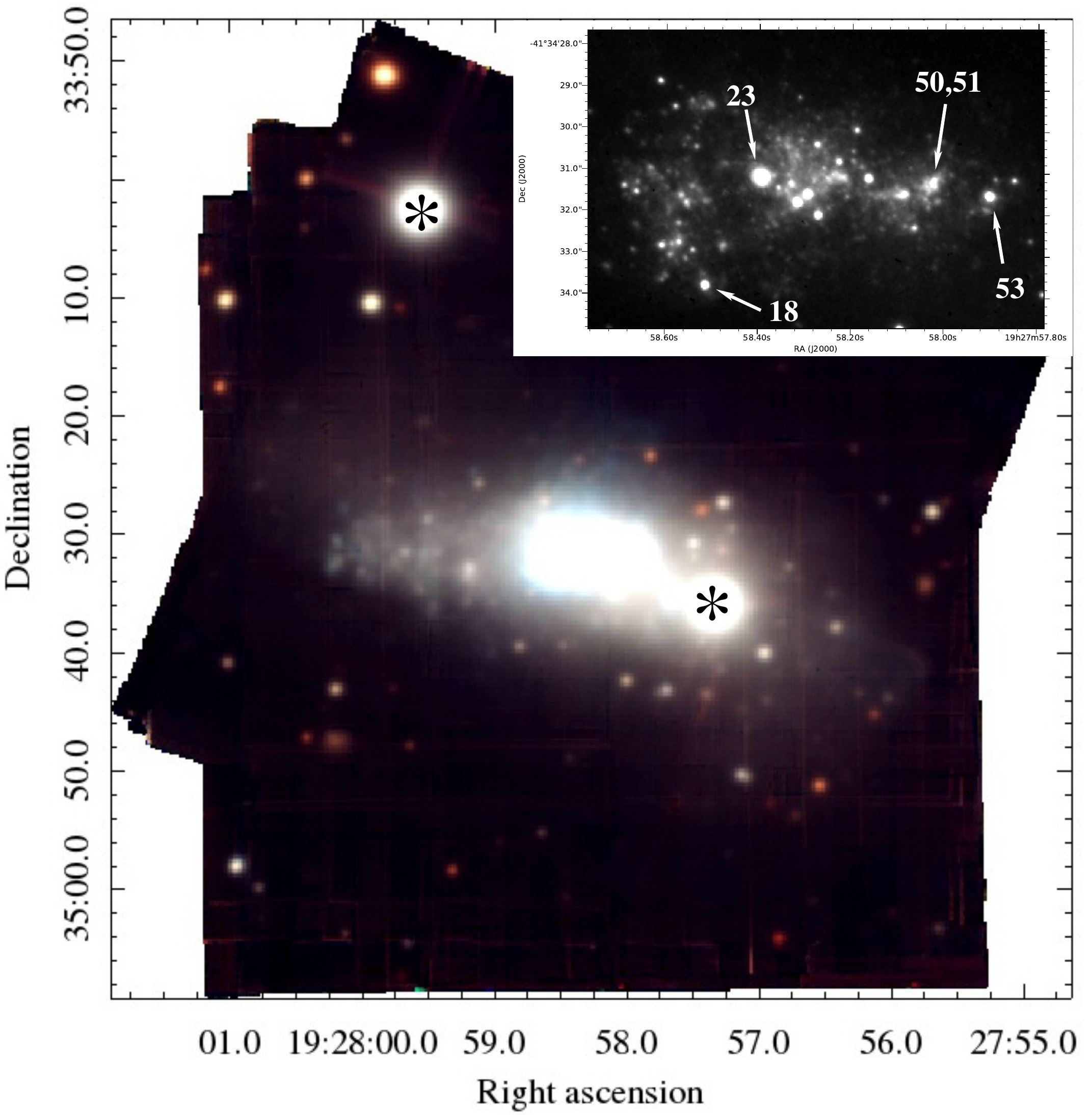}
      \caption{Three colour image made from \emph{VRI} broadband images reconstructed from the MUSE data-cube. The galaxy as well as several star clusters are apparent. The two brightest point sources in the image are foreground stars and are marked with a star symbol. The inset is the V-band image observed with HST \citep{Ostlin09} with marked the 4 clusters with WR features in their spectra.}
         \label{fig:colorimage}
   \end{figure}

\section{Observations and data reduction}\label{sec:observations}

In this paper we present and analyse two datasets obtained with  the integral field spectrograph MUSE \citep{Bacon10}, mounted on the VLT on Paranal, Chile. The first MUSE dataset was obtained during the first science verification run on  2014, June 25. The data were taken in non-AO mode with the extended wavelength setting, providing spectra between 4600 and 9300 \AA\ over a 1 $\square \arcmin$ field of view with a 0.2\arcsec\ pixel scale. The first results of this dataset have been published in \citet{Bik15}.  The integrations were split over 2 observing blocks (OBs) with each four frames of 750 s rotated by 90 $\deg$ each to minimise the systematics. 

The second dataset was obtained during the second commissioning run on 2014, July 30 and 2014, August 3, also in non-AO mode and with the extended wavelength setting. This data was spit over 2 OBs with 4 frames of 900 sec in the first OB and 3 frames of 900 sec in the second OB
(Table \ref{tab:obslog}). No separate sky frames were taken for the datasets.

Both datasets are reduced with the ESO pipeline version 1.2.0 \citep{Weilbacher12}, resulting in an improved data cube, with significantly better sky subtraction than the data presented in \citet{Bik15}, processed with a much earlier version of the pipeline. For the science verification data set,  slit 6 of IFU 10 is vignetted and contains very low flux due to data acquisition at low instrument temperature. The ESO provided trace tables for this slit have been used to get a wavelength calibration. The slit has been removed from the science data. The dataset taken during commissioning did not have this problem.

\begin{table}
\caption{Observing log}
\label{tab:obslog}
\centering
\begin{tabular}{l r c l}
\hline\hline
  pID & Date &  Exp time & seeing (\arcsec)  \\
\hline 
60.A-9314  & 2014-06-25 & 4 $\times$ 750s & 0.77 - 0.99  \\
60.A-9314  & 2014-06-25	&  4 $\times$ 750s &0.67 - 1.01\\
60.A-9100(C) &2014-07-30	&  4 $\times$ 900s &0.82 - 0.92\\
60.A-9100(C) & 2014-08-03 &  4 $\times$ 900s& 1.04 - 1.69\\
\hline
\end{tabular}
\end{table}

\subsection{Sky subtraction}\label{sec:skysub}

As the \halpha\ and  [\ion{O}{iii}]5007\AA\ lines extend to the edge of the observed field of view, special care has been taken to properly remove the sky background. The edges of the fields, where the \halpha\ and  [\ion{O}{iii}]5007\AA\ are not present are too small, resulting in a low-signal to noise sky spectrum. Applying this spectrum results in a bad correction of the emission lines, especially in the red part of the spectrum. 

We therefore apply a different approach to remove the sky emission without  affecting the nebular emission lines. This process is done in two steps. Firstly, a sky background is created with the pipeline using a {\it skymodel\_fraction} = 0.2 in the {\it scipost} procedure. After inspection of the resulting sky spectrum, emission lines of \halpha, \hbeta\,  [\ion{O}{iii}]5007\AA\  and 4959\AA\ are present. These are removed by interpolating below these lines in the {\it SKY\_CONTINUUM} file. The {\it SKY\_CONTINUUM} should contain only a smoothly varying continuum, as it is created after the sky lines have been subtracted.  Any residuals in this spectrum are either sky line subtraction residuals or extended emission lines. None of the selected emission lines were contaminated by sky line residuals. The modified  {\it SKY\_CONTINUUM}  file is then again inserted in the pipeline and {\it scipost} is run again with the modified sky as input, resulting in a much better removal of the sky lines. This approach has been also successfully applied in \citet{Herenz17}. 

We then extract line maps for the emission lines of scientific interest by numerically integrating under the line profile. The continuum underneath the line  is estimated by averaging the continuum left and right of the emission line and subtracted. The corresponding noise frames are extracted from the error cube by quadratically summing the error over the wavelength range the line is integrated. On the line maps we check for residual sky emission or absorption the following way:  We apply a weighted Voronoi tessellation binning algorithm by \citet{Diehl06}, which is a generalization of the \citet{Cappellari03} algorithm, with a signal to noise of 5 and a maximum cell size of 2500 pixels to detect the line emission and subtract the emission with a  signal-to-noise higher  than 5 from the line map. The resulting map now only contains residual sky emission. We fit a Gaussian function to the histogram of the sky values to calculate the centroid of the distribution. The centroid is then subtracted from the original map, making sure that the sky pixels have values centred around 0. The typical sky residuals found this way are below $1.5 \times 10^{-20}$ erg\ s$^{-1}$ cm$^{-2}$.

Finally, the data of the two datasets are combined into one final data cube. This data cube is about twice as deep in exposure time as the data presented in \citet{Bik15}.  The final image quality is 0.9\arcsec as measured on a reconstructed V-band image.
 Fig \ref{fig:colorimage} shows a 3 colour image of the VRI broad-band images extracted from the final data cube, showing the galaxy and the surrounding star clusters as well as some background galaxies. Two bright foreground stars are marked with a star sign.

\section{Young Stellar Clusters}\label{sec:clusters}

ESO 338 hosts a large young stellar cluster (YSC) population, formed during the current starburst. \citet{Ostlin98,Ostlin03} carried out a detailed analysis of the stellar population based on NUV, UBVI  HST imaging. They found that the present starburst has been active since 40 Myrs. \citet{Adamo11} show that the cluster formation rate in ESO 338 has strongly increased since $\sim$ 20 Myrs. They also show that the amount of stellar mass forming in clusters with respect to the total stellar mass formed in ESO 338 is very high ($50 \pm 10 \%$), making this galaxy an excellent show case in studying the feedback of YSCs on the ISM. 

  \begin{figure*}[!t]
   \centering
   \includegraphics[width=\hsize]{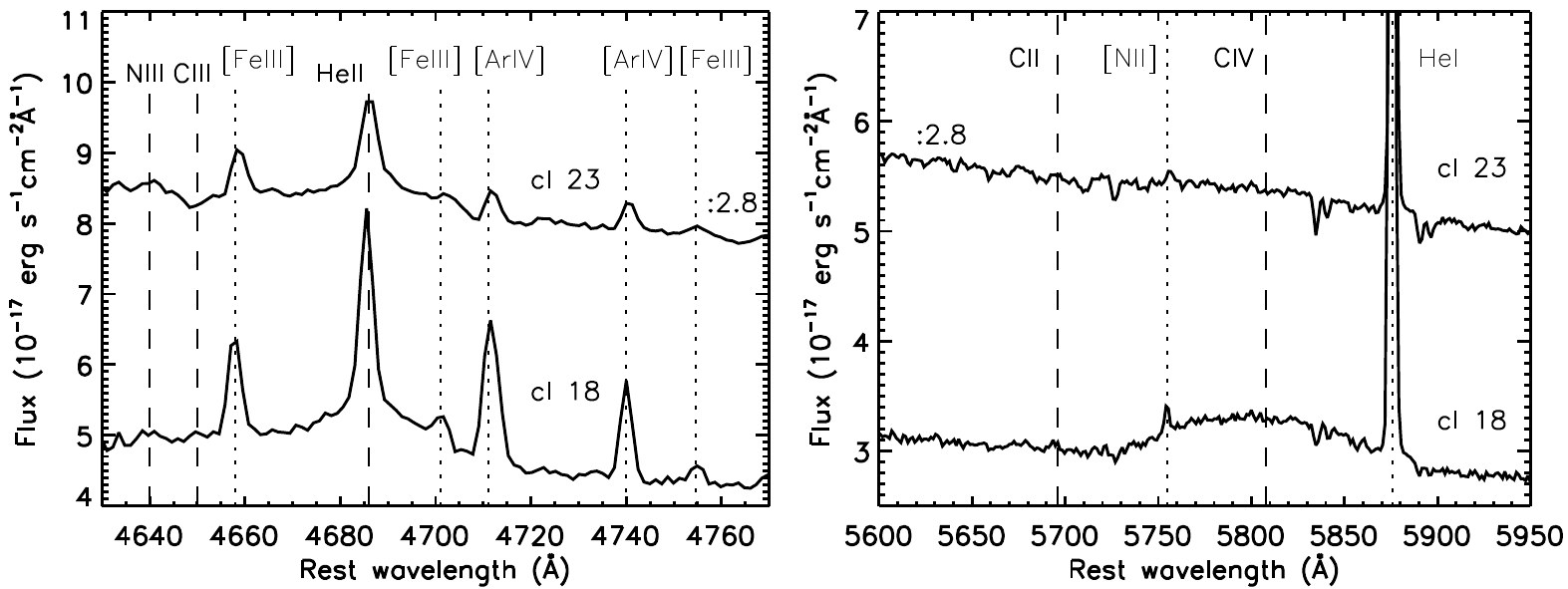}
      \caption{Spectra of cluster 23 (top) and cluster 18 (bottom) highlighting the WR features in their spectra. The spectrum of cluster 23 is scaled down with a factor 2.8 to fit in the same scale. \emph{Left:} The spectral region around the \ion{He}{ii} 4686\AA\ line. The broad emission component is the signature of the presence of WR stars, the other WR lines (\ion{N}{III} and \ion{C}{III}) are not detected. The narrow component of \ion{He}{ii} is originating in the diffuse gas. Additionally, all the nebular emission lines are marked. \emph{Right:} The spectral region around the red WR bump. This emission bump is originated by \ion{C}{IV}, tracing WC stars. Cluster 23 only shows a faint bump, while cluster 18 shows a very strong feature.} 
         \label{fig:WRfeature}
   \end{figure*}

\citet{Ostlin09} present additional \halpha\ imaging as well as far-UV data, allowing a better constraint on the ages of the very young clusters, covering the energy distribution between $\sim$1400\AA\ to 9500\AA, including narrow band \halpha. 
Following the procedure outlined in \citet{Adamo10}  we construct spectral energy distributions (SEDs) of the clusters.  A cluster catalogue is created by detecting the clusters in the F550W and  F814W filters independently using SExtractor \citep{Bertin96}. 
The final cluster catalogue is created by merging the catalogues of the two filters and requiring a detection in both the F550W and  F814W images.  Aperture photometry was performed to all bands with the cluster catalogue as input with fixed aperture size of 0.125\arcsec\ in all the bands. The inner radius of the sky annulus was set at 0.15\arcsec\ with a width of 0.05\arcsec.  

The bright clusters discussed in this paper have a high signal to noise detection in all  6 bands (UV (F140LP), U (F336W), B (F439W), V (F550M), \halpha\, (FR656N), I (F814W) and their SEDs are fitted to the latest version of the Yggdrasil models \citep{Zackrisson11} using $\chi^2$ minimization. These models take into account the contribution of both single stellar population models \citep[Starburst99,][]{Leitherer99} with a \citet{Kroupa01} IMF and gaseous continuum and emission lines produced with Cloudy \citep{Ferland13}.

The models assume a stellar metallicity of Z=0.004 and a density of the ionized gas of $n_e$ = 100 cm$^{-3}$ \citep[See for more details][]{Adamo17}. The model fits are performed for three different covering fractions (0, 0.5 and 1.0). A covering fraction of unity means that all the LyC photons emitted by the cluster are absorbed by  surrounding gas inside the photometry aperture, resulting in nebular emission.  As shown in \citet{Adamo10SBS} the strongest source of nebular contribution is the free-free emission. For the first 3-4 Myrs, this can be as high as 50-60\% of the flux in the broadband filters, but strongly declines after 4 Myrs.

The overall results of the fit are similar to the results presented in \citet{Ostlin98,Ostlin03}, showing that ESO 338 contains numerous YSCs with ages between 0 and 40 Myrs, whose ionizing photons and mechanical energy strongly affect the surrounding ISM. We will discuss the stellar population in detail in a forthcoming study and  for this paper we only look at some of the most massive young stellar clusters where we find evidence for Wolf-Rayet (WR) stars based on their MUSE spectra.

  \begin{figure*}[!t]
   \centering
   \includegraphics[width=0.98\hsize]{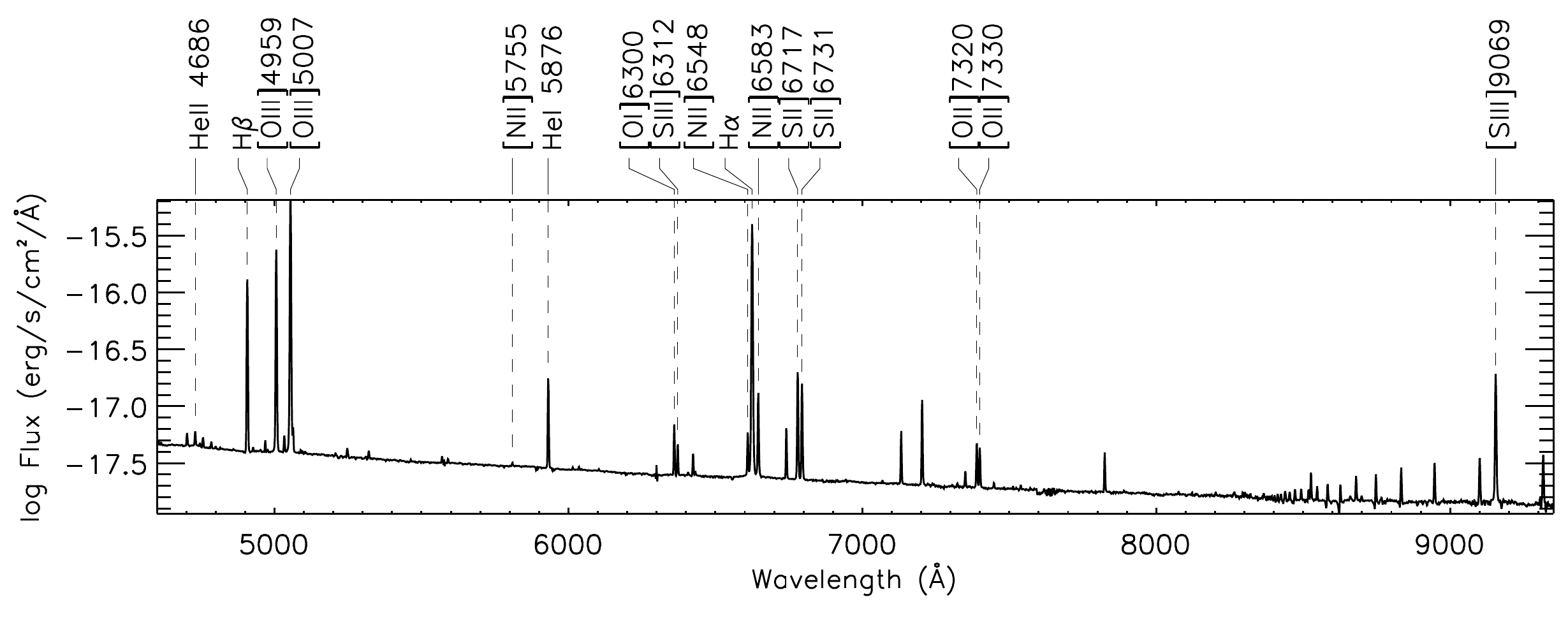}
      \caption{Integrated MUSE spectrum centred on cluster 23 ($\alpha$ =19$^{\rm{h}}$27$^{\rm{m}}$58.429$^{\rm{s}}$, $\delta$=-41\degr34\arcmin 30.74\arcsec) extracted with an aperture of 3\arcsec, covering the central regions of ESO 338. The emission lines used in this paper are labeled in the spectrum. Note: this spectrum contains one of areas where \oiii\ and \halpha\ are saturated.}
         \label{fig:spectrum}
   \end{figure*}

\subsection{Wolf-Rayet clusters}\label{sec:wrclusters}

The WR phase is a short phase in the life time of a massive star. In this evolutionary stage, the evolved massive stars expel their outer layers with a very strong stellar wind \citep{Crowther07}. As the inner regions of the stars become optically  visible, their effective temperatures can be as high as 100,000 K \citep{Smith02}.  This hot temperature makes the WR stars important contributors to the emitted He$^{+}$ ionizing photons.  Additionally, due to their strong winds, they also contribute to the chemical enrichment of the galaxies ISM \citep[e.g.][]{MonrealIbero12}.
However, at low metallicity the contribution of the WR stars  in the ionizing budget becomes less important \citep{Kunth86,Guseva00} and O stars significantly contribute to the He$^{+}$ ionizing budget  \citep{Brinchmann08}.   Recent studies show that standard population synthesis models cannot account for the amount of observed \heii\ emission \citep{Kehrig15,Kehrig18}. After excluding possible contributions of shocks and gas accretion,  very massive stars, or very low metallicity stars need to be used to explain the observed \heii\ emission.

Assuming a single burst of star formation, massive stellar clusters with ages between  3 and 5 Myrs show spectral features originating from WR stars  \citep[e.g.][]{Leitherer99,Smith02}. The spectral features of WR stars in the optical consist of two broad emission features (Fig. \ref{fig:WRfeature}). The "blue" WR bump consists of broad  \ion{He}{II} 4686 \AA\ emission as well as fainter \ion{C}{III} and \ion{N}{III} emission, while the "red" WR bump is observed at 5808\AA\ and is caused by \ion{C}{IV} emission \citep[e.g.][]{Schaerer99,Smith16}. 

In order to find the clusters with these WR features we construct a pseudo narrow band image (width = 100 \AA) covering the red WR bump and subtracted the underlying continuum by interpolating between the blue and the red side of the emission band.  We use the same method to construct a pseudo-image of the broad \ion{He}{ii} emission. Here we  select the broad emission only and leave out the narrow \ion{He}{ii} emission as well as [\ion{Fe}{III}] and [\ion{Ar}{IV}] with a width of 10 \AA\ left and right of the narrow  \ion{He}{ii} emission. The velocity differences of the ionized gas are relatively low (see Sect. \ref{sec:kinematics}), therefore there is no contamination expected from the [\ion{Fe}{III}] and [\ion{Ar}{IV}] in the pseudo images. The narrow  \ion{He}{ii}  is originating in the diffuse gas and will be discussed in Sect. \ref{sec:ionization}.

\begin{table}[!ht]
\caption{Clusters in ESO338 with WR features}
\label{tab:wrclusters}
\centering
\begin{tabular}{r|ccrcc}
\hline\hline
 ID\tablefootmark{a} & Age & Mass & Cov. & blue & red \\
& (Myrs) & ($10^5$ M$\sun$) & fraction & bump & bump\\
\hline
23	& 4  &38  & 0.0 & Y & Y\\
18   & 3 -- 6   &2.4 -- 4.5 & 0.0 -- 0.5& Y & Y\\
50/51\tablefootmark{b}  & 3 -- 4  &  0.8 -- 1.3  & 0.5 -- 1.0& Y & Y\\
53 & 3 -- 4 & 3.6 -- 4.0 & 0.5 -- 1.0& Y & N\\
\hline 
\hline
\end{tabular}
\tablefoot{
\tablefoottext{a}{IDs taken from \citet{Ostlin98}}
\tablefoottext{b}{This source was resolved by \citet{Ostlin98} in two clusters, while with the new fitting, it was selected as 1 extended source}
}
\end{table}

We find 4 clusters with WR features in their spectrum; all four show the broad \ion{He}{ii} emission and three clusters show the red WR bump. These clusters show clear signs of the WR features and are rather isolated such that the emission can unambiguously be related to the cluster. Fainter clusters might well have WR features present, but they were not visible as strong signals in the narrow-band images, and also are more difficult to detect due to the lower spatial resolution of MUSE compared to HST.
The clusters are identified on the HST V-band image in Fig. \ref{fig:colorimage} using the IDs from the inner sample of \citet{Ostlin98}. Clusters 18, 23 and the source coinciding with clusters 50 and 51 all show both the blue and the red WR features, while cluster  53 only shows the broad \ion{He}{ii} emission. Fig. \ref{fig:WRfeature}  shows the blue  and red  WR features of cluster 23  and 18 extracted with a circular aperture with a 2 pixel (0.4\arcsec) radius.  This aperture is small enough that the continuum emission from the clusters could be isolated and is free from contamination of other bright clusters. 
We did not subtract any background as in the continuum that is negligible (they are very bright clusters) and it would only create over- or under-subtraction of the nebular lines as they are varying in intensity on small spatial scales. Both clusters show the broad \ion{He}{ii} emission feature at 4686 \AA\ with super imposed a narrow component originating in the ionized \ion{H}{ii} region. Also other nebular emission line from [\ion{Ar}{IV}] and [\ion{Fe}{III}] are identified. From the two other WR lines (\ion{N}{III} and \ion{C}{iii}) we see a hint of \ion{N}{III} emission in cluster 18.

Table \ref{tab:wrclusters} summarizes the properties of the clusters as derived from fitting the  Yggdrasil models to the cluster SEDs. All 4 clusters are among the most massive in the galaxy with masses of $\sim$$10^5$ \msun\ or higher. The reason why only those clusters show detected WR features is that they have the highest signal-to-noise spectra. There will be many more lower-mass clusters containing WR stars, but their features will be fainter and harder to detect. Below we discuss the properties of each cluster:

\emph{Cluster 23:}  is the most massive cluster of ESO 338 with a photometric mass of $3.8 \times 10^6$ \msun. \citet{Ostlin07} derive a dynamical mass based on cross correlation of the spectrum with stellar templates and derived a dynamical mass of $1.3 \times 10^7$ \msun. The discrepancy between the photometric and dynamical mass can be explained by the presence of spectroscopic binaries \citep{Portegies10}.  From fitting the Balmer absorption lines,  they also derived an age of $6^{+4}_{-2}$ Myr. This cluster is located inside a large bubble which is the results of evacuated gas due to the  feedback of the stellar population inside cluster 23 \citep{Ostlin09}.  Based on the expansion speed of \oiii\ and \halpha\ an expansion age of the bubble of 3--4.5 Myr is derived  \citep{Ostlin07}, suggesting that the cluster expelled its gas very early in its evolution \citep{Bastian14}. 

By modelling the feedback from stellar winds and supernova explosions \citet{Krause16} model the shell expansion around cluster 23 and try to explain the size and velocity of the bubble. By assuming a star formation efficiency of 30\%, the energy injected by stellar winds and supernovae is not enough to clear the gas out of the cluster and create a bubble of the size observed. A star formation efficiency as high as 80\% is required to explain the rapid gas expulsion of cluster 23. 

The age of the cluster as result of our SED fitting strongly depends on the choice of covering fraction. For cluster 23, a covering fraction of 0.0 was chosen, as the gas is blown out well beyond the photometry aperture of 0.125\arcsec. The derived age (4 Myr) and mass (3.8$\times 10^6$ \msun)  (Tab. \ref{tab:wrclusters}) are consistent with previous measurements as well as with the presence of WR features in the spectrum (Tab. \ref{tab:wrclusters}).  If we would, unrealistically, assume a much higher covering fraction, the low intensity of \halpha\ close to the cluster would have resulted in a much older age; for a covering fraction of 1.0 the SED fitting produces an age of 15 Myrs.

\emph{Cluster 18:} is located in the south-eastern part of the galaxy and also has cleared out most of it's gas content. It is located near a bow-shock like feature visible in \halpha\ \citep{Ostlin09}. The ages and masses derived for this cluster assuming a covering fraction between 0 and 0.5 are 3 -- 6 Myrs and $3.6 - 4.0 \times 10^5$ \msun\ respectively. Similar to cluster 23, in the case of a unrealistically covering fraction of 1.0, the derived age would be 15 Myrs.
This cluster has the strongest red WR bump of all 4 clusters, clearly identifying the presence of WC stars.

  \begin{figure*}[!t]
   \centering
   \includegraphics[width=0.98\hsize]{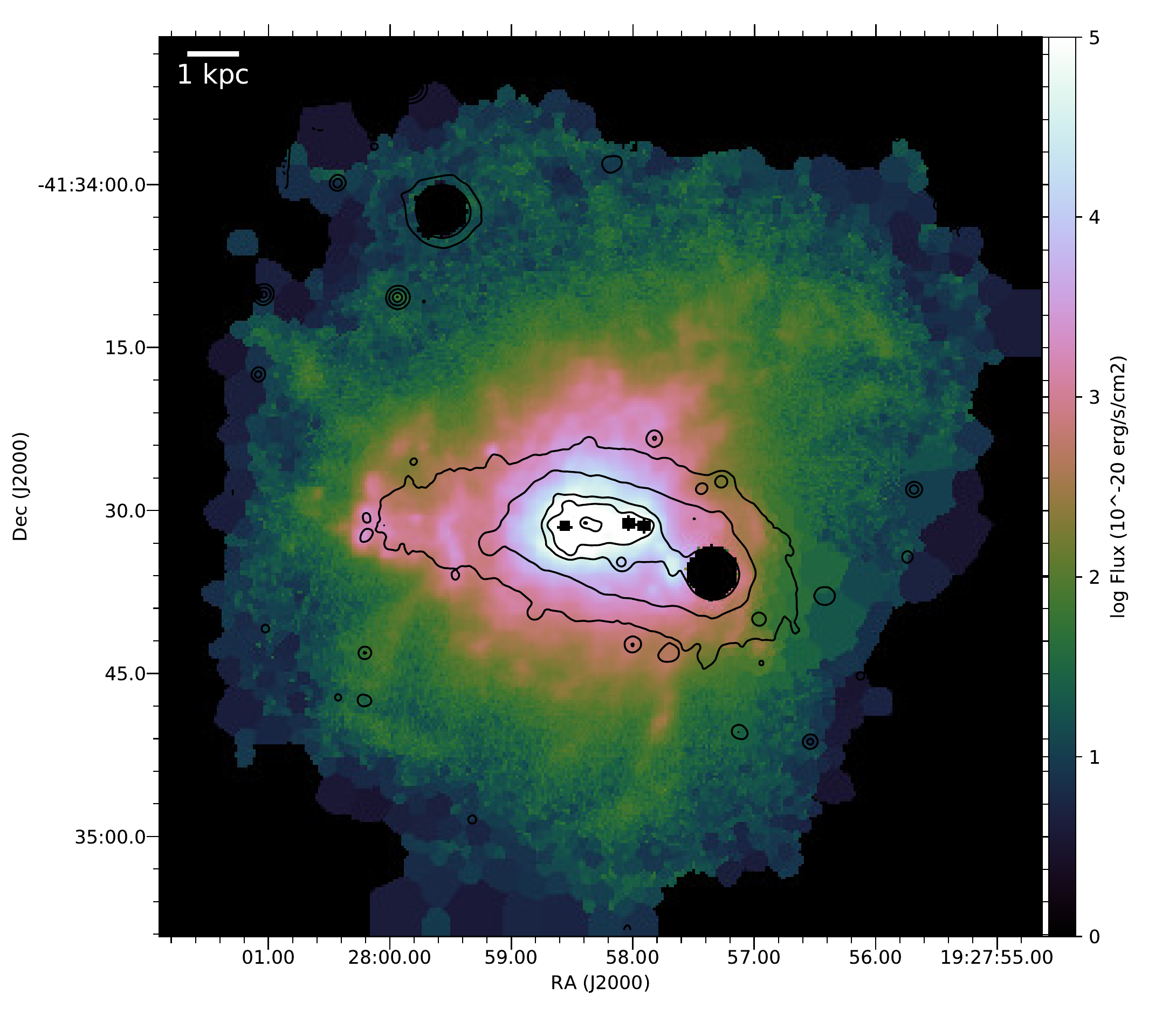}
      \caption{Logarithmic scaled \halpha\ emission line map of ESO 338 showing the spatial extend of the ionized halo. A Voronoi binning with minimum signal-to-noise of 5 and maximum bin size of 900 pixels ($\sim$6$\square$$\arcsec$) was applied. The Voronoi cells which have lower signal-to-noise than 5 are removed and plotted as black.  Plotted in black contours are the contours of the I-band  image reconstructed from the MUSE cube. Contour levels are 15, 50, 100, 400, 700 and 1000  $\times$ $10^{-20}$ erg/s/cm$^2/\AA$. The lowest contour corresponds to an AB I-band magnitude of 25.2 mag. The saturated parts of the H$\alpha$ image in the center as well as the two bright foreground stars are masked out and appear as black.}
         \label{fig:halpha}
   \end{figure*}

\emph{Clusters 50/51} were identified by \citet{Ostlin98} as two separate clusters.  However, in our analysis of the HST data, only a single source has been identified. This cluster has strong \halpha\ emission and has most likely not cleared its gas yet. Therefore we have chosen a covering fraction between 0.5 and 1.0. It is the lowest mass system in our sample with a mass of 0.8 -- 1.3 $\times 10^5$ \msun. 

\emph{Cluster 53} show a weak blue WR bump and no red bump, suggesting that not many WR are present. Also this cluster has likely not cleared it's gas content yet as a lot of \halpha\ emission is present on-source, favouring a high covering fraction.

\section{Physical properties of the ionized gas}\label{sec:gasprop}

In this section we investigate the properties of the ionized gas in the halo of ESO 338. We derive physical properties like the extinction, density,  and level of ionization using the recombination and collisionally excited lines  present in the MUSE spectrum of ESO 338 (Fig. \ref{fig:spectrum}). Additionally, we discuss the kinematics of the halo based on the \halpha\ emission line.  

Fig. \ref{fig:spectrum} shows the MUSE spectrum integrated over  a circular aperture with a radius of 3 \arcsec, centred on the position of cluster 23 with the most important lines annotated. The H$\alpha$ and the [\ion{O}{III}] 5007 \AA\ lines are the strongest lines in the data cube. Towards the three brightest regions in the galaxy these lines appear saturated in a large fraction of the individual exposures. All the other emission lines are significantly fainter and saturation is no issue.  In the analysis of the emission lines where H$\alpha$ and/or [\ion{O}{III}] 5007\AA\ are included, we do not use the saturated parts of the data and they are clearly marked in the H$\alpha$ line map (Fig. \ref{fig:halpha}) and other figures where these lines are used.

\begin{table*}[!ht]
\caption{Diagnostics}
\label{tab:diag}  
\centering
\begin{tabular}{l|r|l|l|r|r|r}\hline\hline
Diagnostic & Ion & Lines  &\multicolumn{2}{c|}{Voronoi binning  }& Ref. & Figure\\
		& 	&  (\AA)		& S/N & max. area& &\\
		& 	&  		& & (sq. pixels)& &\\

\hline
Size		    & H & {\bf \halpha} & 5 & 900 & & \ref{fig:halpha},\ref{fig:sb_halpha} \\
Dynamics     & H & {\bf \halpha} &20 & 900  & & \ref{fig:kinematics}\\
Extinction	     & H & \halpha, {\bf \hbeta} & 30 & 900 & 1 & \ref{fig:ebminv}\\
Density	     &[\ion{S}{II}] & 6716, 6731 &--- & --- &2  & \ref{fig:ne_sii}\\

Ionization 	     &[\ion{S}{II}], [\ion{O}{III}] & {\bf 6716+6731}, 5007 & 10 & 900& 3& \ref{fig:ionization}\\
	             &H, [\ion{O}{III}] & H$\alpha$, {\bf 5007}	 & 10  & 900& & \ref{fig:ionization} \\
WR stars & \ion{He}{II} & 4686	&      ---       & --- & & \ref{fig:ionization}\\
BPT		     & [\ion{O}{III}], H, [\ion{S}{II}] &5007, \halpha, \hbeta, {\bf 6717 +6731} & 20 & 900 &4 &\ref{fig:bpt}, \ref{fig:bpt_sel}  \\
		     & [\ion{O}{III}], H, [\ion{N}{II}] &5007, \halpha, \hbeta, {\bf 6583}  & 20 & 900 &5&\ref{fig:bpt}, \ref{fig:bpt_sel}   \\
		     & [\ion{O}{III}], H, [\ion{O}{I}] &5007, \halpha, \hbeta, {\bf 6300} & 20 & 900 &5  &\ref{fig:bpt}, \ref{fig:bpt_sel} \\

\hline
\end{tabular}
\tablefoot{The lines marked in boldface are used to calculate the Voronoi pattern. For the density we only calculate the radial profile, where no Voronoi binning is applied.}
\tablebib{(1)~\citet{Prevot84}; (2)~\citet{Luridiana15}; (3)~\citet{Pellegrini12}; (4)~\citet{Baldwin81}; (5)~\citet{Veilleux87}}
\end{table*}

Table \ref{tab:diag}  lists the emission lines used for each diagnostic. We apply  Voronoi binning to all diagnostics in order to increase the signal to noise in the outer regions of the halo. The Voronoi pattern has been calculated on the faintest line (marked in boldface) and is then applied to the brighter line(s) used for the same diagnostic.
For the diagnostics where line ratios are calculated we typically choose a higher signal to noise than for just emission lines maps.

\subsection{Correction for stellar absorption}\label{sec:hbeta}

Two of the important emission lines used in the analysis of the ionized gas in ESO 338 are  \hbeta\ (4861\AA) and \halpha\ (6563\AA). Inspection of the \hbeta\ line profile shows absorption underneath the emission line in some locations of the galaxy. This is caused by a young and intermediate age underlying stellar population \citep{GonzalezDelgadoI99,GonzalezDelgadoII99}. This absorption will also be present underneath \halpha, but much harder to see due to the brightness of \halpha\ and nearby [\ion{N}{ii}] lines. In order to measure the correct strength of the emission lines, this absorption needs to be corrected for. We correct for the absorption lines by fitting simultaneously the emission and underlying absorption. In the regions of the galaxy where stellar continuum is present we fit the emission with a Gaussian profile and the absorption with a Lorentzian profile, better reproducing the broad absorption winds of the \hbeta\ absorption.

To determine  where to estimate the continuum level, we smooth the cube with a boxcar with a width of 5x5 pixels (1\arcsec$^2$). Following  \citet{GonzalezDelgadoI99}, we measure the continuum near  \hbeta\  at $\pm$ 30 $\AA$ away from the line center with a width of 10 $\AA$ for each side. A linear fit between the two continuum apertures was performed and removed from the  \hbeta\ spectrum. From these background values a continuum image was constructed.  We choose a flux level of  $3.0 \times 10^{-19}\ \mbox{erg}\ \mbox{s}^{-1}\ \mbox{cm}^{-1}\ \AA^{-1}$  per pixel to determine whether a continuum is present or not.

For the regions where continuum is present both a fit with a single Gaussian emission and a fit with  a Lorentzian absorption and Gaussian emission profile are being made. The solution with two lines is only selected if the $\chi^2$ more than 10\% better than that of the single line fit. The fit is performed on the smoothed data cube to increase the area of the galaxy where the absorption can be fitted. The correction on the emission line is calculated from the fit to the smoothed data and applied to the un-smoothed emission line flux.   The regions in the cube without continuum  do not need to be corrected and are therefore fit with a single Gaussian tracing the emission line. Finally, \halpha\ is corrected for the underlying absorption, adopting a ratio between the EW of \halpha\ and  \hbeta\   of 0.65 as derived by \citet{Olofsson95} for a Z=0.001, young ($<$ 15 Myr), stellar  population. This value varies with age and initial mass function (IMF)  of the underlying stellar population. As the applied corrections are typically very small, this has negligible influence on the corrected \halpha\ emission line flux.
The applied corrections to the  \hbeta\  line-flux map range from less than 1\%  in the central 10\arcsec\ of ESO 338, where the absorption cannot be detected due to the very strong emission to 60\% towards some of the older clusters in the outskirts of the halo.

Fig. \ref{fig:halpha} shows the absorption corrected emission line map of H$\alpha$. A Voronoi binning with a minimal SNR of 5 and a maximum area of 900 pixels was applied and result in a detection limit for \halpha\ of  $6.25 \times 10^{-19}$  erg s$^{-1}$ cm$^{-2}$ $\square \arcsec$,  integrated over the line profile.

In \citet{Bik15} the H$\alpha$ line map already has been presented, showing the large halo which is surrounding ESO 338. However, with the inclusion of the commissioning dataset the depth of the dataset is twice as deep in exposure time and reveals the presence of ionized gas even further out to a distance of $\sim$8 kpc. Additionally, the halo is not smooth as a lot of spatial structure is visible in the emission map. Towards the South some filaments pointing outwards and to the North a more clumpy structure are present.  Towards the northern edge the \halpha\ emission seems to be even more extended than the observed frame.

  \begin{figure}[!t]
   \centering
   \includegraphics[width=0.98\hsize]{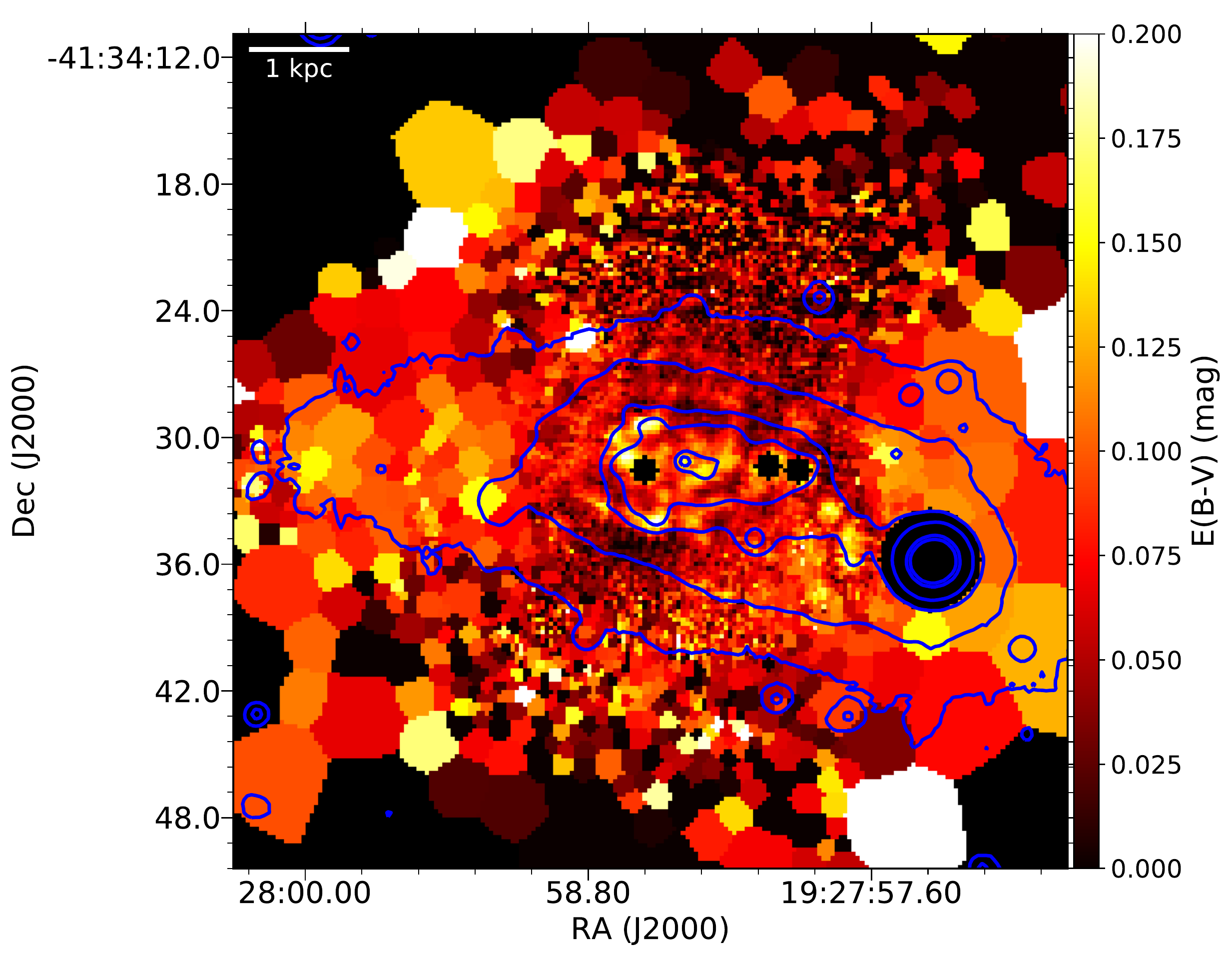}
      \caption{ E(B-V) map towards ESO 338 derived from the \halpha/\hbeta\ ratio. Both lines are corrected for underlying absorption of the stellar population. The extinction law of \citet{Prevot84} is used and the plotted E(B-V) is not corrected for the galactic foreground extinction. The blue contours are the  contours of the I-band  image reconstructed from the MUSE cube (identical to the black contours in Fig. \ref{fig:halpha}).   The saturated parts of the H$\alpha$ image in the center as well as the two bright foreground stars are masked out and appear as black. The high E(B-V) values at the outskirts are cells with relatively low signal-to-noise. The figure is  zoomed in with respect of Fig \ref{fig:halpha}.}
         \label{fig:ebminv}
   \end{figure}

\subsection{Extinction}\label{sec:ebminv}

We use the ratio of the stellar absorption corrected \halpha\ and \hbeta\ to calculate a spatially resolved extinction map. As the metallicity of ESO 338 is low \citep{Ostlin03,Bergvall85} and similar to the SMC, we apply the extinction law of \citet{Prevot84} to determine the E(B-V) of each pixel.  
The extinction coefficients we take from \citet{McCall04}.  We adopt an intrinsic value of  \halpha\ over \hbeta\ ratio of 2.86, assuming Case B, a density of 10$^2$ cm$^{-3}$ and an electron temperature of 10$^4$ K \citep{Osterbrockbook}.
We calculate the Voronoi tessellation pattern on the \hbeta\ map to make sure that the individual cells  have  a minimum signal-to-noise of 30 over a maximum area of 900 pixels (6 $\square$$\arcsec$), aiming at an error of $\sim$0.04 mag in E(B-V). This pattern is applied to the \halpha\ line map and the ratio map is made after that.

Fig. \ref{fig:ebminv} shows the resulting spatially resolved E(B-V) map. The E(B-V) map covers a smaller spatial extend than the \halpha\ emission line map as the \hbeta\ line is fainter and the minimum signal-to-noise for the Voronoi binning is reached at smaller radii. The E(B-V) map is corrected for a galactic foreground extinction of E(B-V)$_{fg}$ = 0.0742 $\pm$ 0.0011 mag \citep{Schlafly11}. The mean value for E(B-V) towards ESO 338 is 0.066 mag, with a standard deviation of 0.077 mag. The typical error  per pixel on the E(B-V)is 0.03 mag, with smaller values in the center where the \halpha\ and \hbeta\ flux are very high and higher values, up to 0.05 mag near the outskirts of the halo. 

This map shows that ESO 338 has very low, or even no, extinction by dust. This is consistent with previous measurements of the extinction towards ESO 338 based on long-slit spectroscopy \citep[e.g.][Rivera-Thorsen et al, submitted]{Bergvall85} or SED modelling of the cluster population \citep{Ostlin98,Ostlin03}.

We can see some spatial variation in the  E(B-V) map. The highest values are reached to the eastern and western part of the galaxy. This is also where the \hbeta\ absorption correction is the strongest. Lower values are reached extending away from the galaxy into the halo in northern and southern direction. This coincides with the  two ionization cones identified in \citet{Bik15} and the low E(B-V) is  consistent with  the high degree of ionization where dust is not expected to survive (see also Sect. \ref{sec:ionization}).  We also find an increase in the measured E(B-V) values towards the HII regions surrounding the massive clusters in the center of the galaxy. The E(B-V) values become as high as 0.18 mag, compared to $\sim$ 0.05 mag in the surrounding gas.  This  indicates that there is still some dust present tracing the left-overs of the giant molecular clouds the clusters have formed in.

  \begin{figure}[!t]
   \centering
   \includegraphics[width=0.98\hsize]{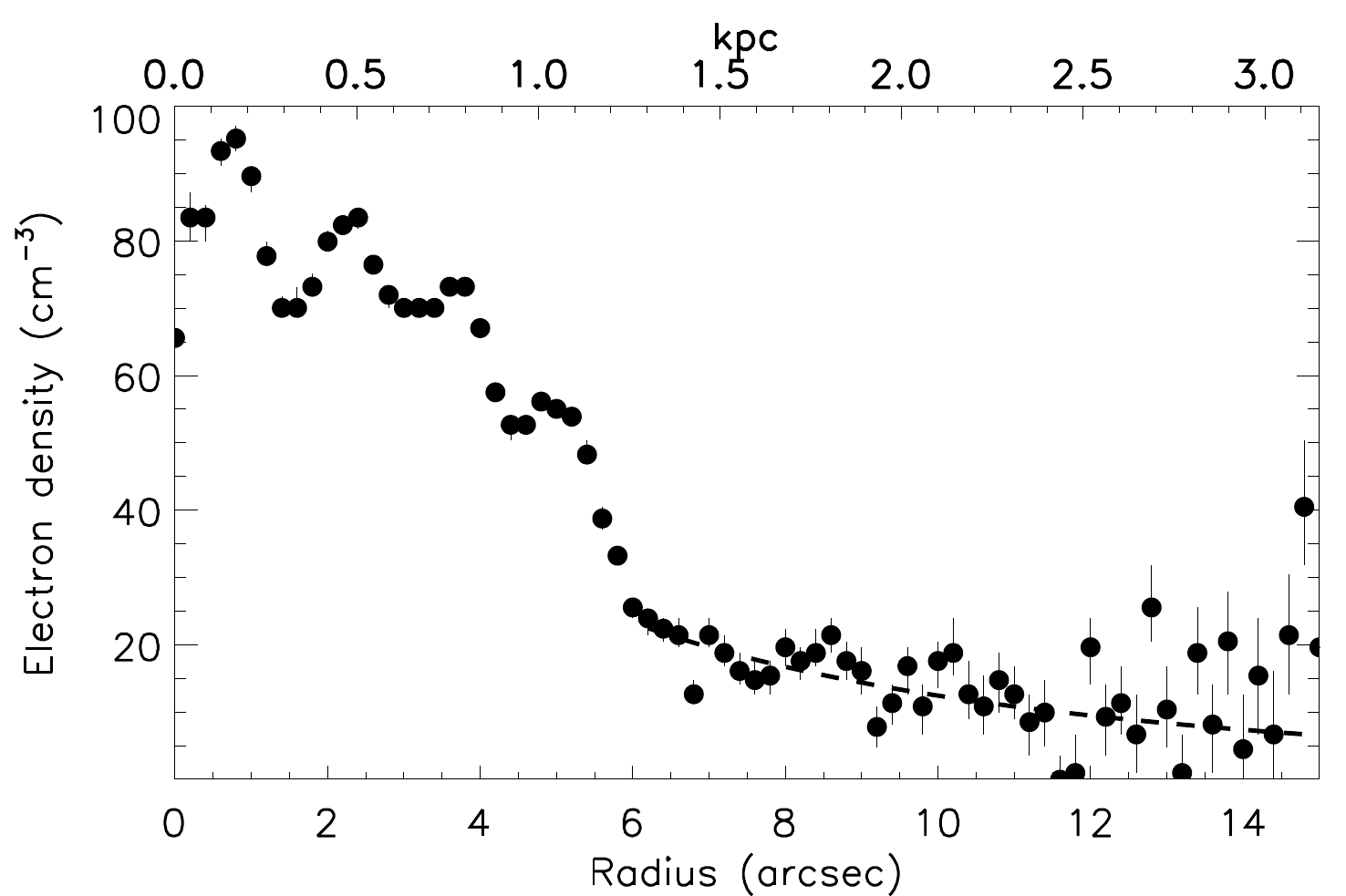}
      \caption{Radial profile of the electron density derived from the ratio of the [\ion{S}{II}] 6717\AA\ and 6731\AA\ lines calculated with pyneb \citep{Luridiana15}.  The dashed line shows the fitted radial profile of the density as explained in Sect. \ref{sec:HIImass}.}
         \label{fig:ne_sii}
   \end{figure}

\subsection{Electron density}\label{sec:tempden}

In the MUSE spectral range we find several line ratios which can be used as a tracer for electron density or temperature. We can derive the electron density from the [\ion{S}{II}] 6717\AA\ / 6731\AA\ line ratio. Two sets of lines can be used to derive the temperature:  [\ion{S}{III}] lines and the [\ion{N}{II}] line ratios. For the temperature, the detection of the faint auroral line ([\ion{N}{II}] 5755\AA, or [\ion{S}{III}] 6312\AA) is required. These lines can only be detected towards the bright central  10x15\arcsec of the galaxy. The lines for the density ([\ion{S}{II}] 6717\AA\ and 6731\AA), however, are much brighter and the density can be traced much further out. The spatially resolved maps of the temperature and density of the inner area of the galaxy will be part of a future study. 
 
In this paper we derive the radial profile of the electron density in order to derive the total mass of the ionized halo (Sect. \ref{sec:halo}).   To derive the electron density as far out in the halo as possible, we construct a radial profile of both  [\ion{S}{II}] emission lines. The center of the radial profile is the location of the brightest, central cluster, cluster 23 (the same positions as the extracted spectrum of Fig. \ref{fig:spectrum}).  The bin width of the radial profile is chosen such that a minimum signal-to-noise of 30 is reached in the faintest [\ion{S}{II}]6731\AA\ line up to a maximum bin width of  2\arcsec.  The electron density is derived by adopting a temperature of 12000 K (a typical value derived from both temperature traces in the central part of the halo) using \emph{getTemDen} in pyneb.

Errors are calculating using a Monte Carlo approach. Based on the derived errors on the line fluxes, the line ratio is calculated 1000 times by randomly drawing from a Gaussian distribution centred around the measured line flux with a $\sigma$ representing the error on the line flux. From each of these line ratios a density is calculated using pyneb. For each radial bin we have a distribution of 1000 values for $n_{e}$ and the 1$\sigma$ errors are derived by taking the 16\% and 84\% values for the lower and upper error respectively.

Fig.  \ref{fig:ne_sii} shows the resulting radial profile. We are able to calculate the density to a radius as far as 15\arcsec (3.2 kpc). Beyond this the derived densities show a scatter which is much larger than the statistical error bars derived from the Monte Carlo simulations. The signal-to-noise of all these points is above the minimum value of 30 set by the construction of the radial profile. Systematic errors, due to, for example, non-perfect background subtraction can be responsible with this. At low densities, the [\ion{S}{II}] line ratio approaches an asymptotic value and small deviations could result in relatively large fluctuations in density. Therefore we do not trust the derived densities beyond the 15\arcsec\ (3.2 kpc) radius. 

In the radial profile, the density drops from a maximum value of around $n_{e} = 100\ \mathrm{cm}^{-3}$ in the center of ESO 338 to values below $n_{e} \approx 10\ \mathrm{cm}^{-3}$ where the densities become difficult to measure as the line ratio approaches the asymptotic value. The central 0.5\arcsec\ shows a lower density due to the bubble evacuated by cluster 23 \citep{Ostlin07}.

  \begin{figure}[!t]
   \centering
   \includegraphics[width=\hsize]{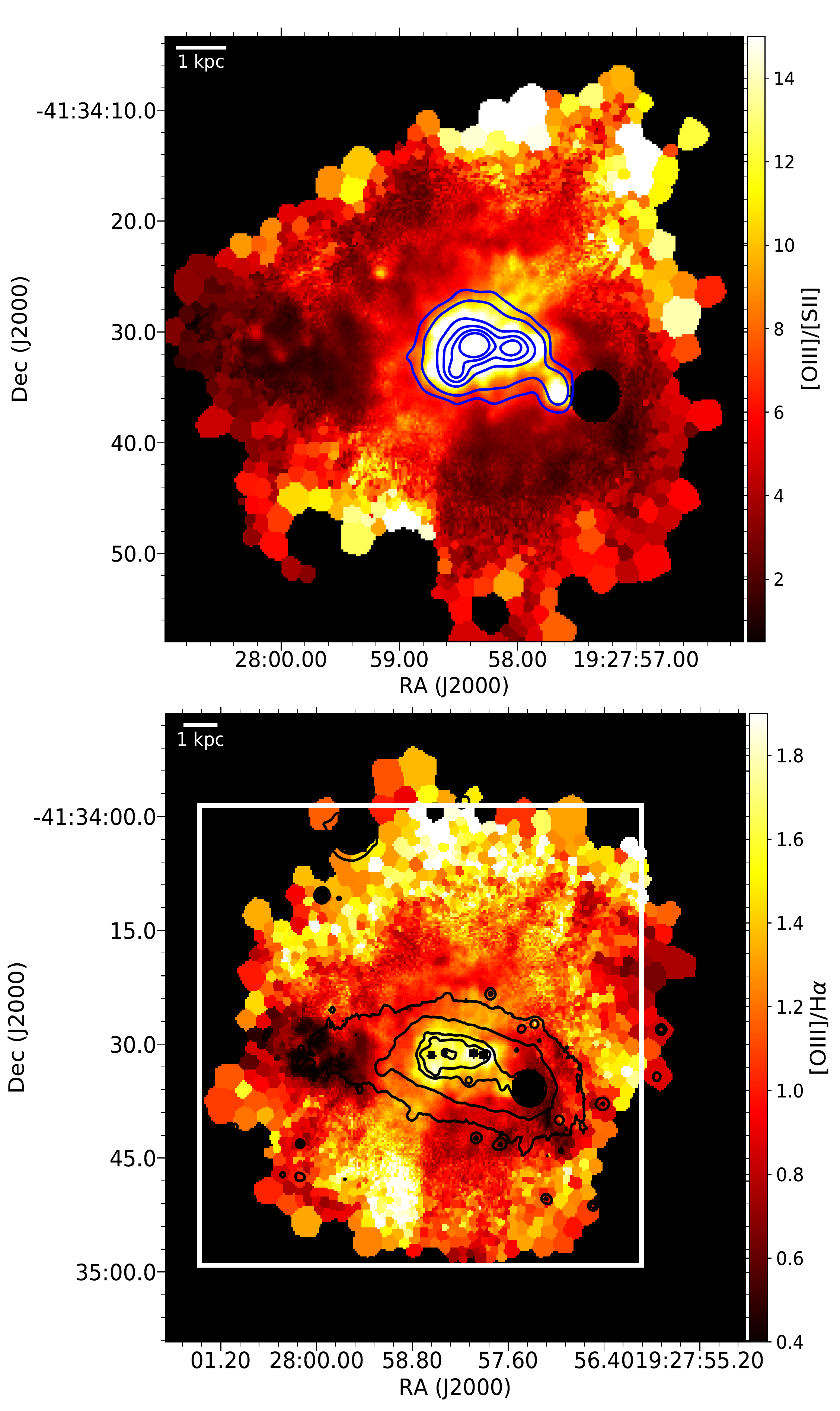}
         \caption{The ionization structure of the halo of ESO 338.       
         \emph{a:} The ionization parameter map of the central area of ESO 338 as traced by the  [\ion{O}{III}]$\lambda$ 5007 $\AA$ / [\ion{S}{II}] $\lambda$(6717+6731)$\AA$  line ratio. The blue contours show the \ion{He}{II} $\lambda$4686\AA\ emission. The \ion{He}{II} image was smoothed with 3 pixel Gaussian kernel. The contour levels range from 50 to 1300 $\times 10^{-20} $ erg s$^{-1}$ cm$^{-2}$. \emph{b:} The ionization map as derived from the [\ion{O}{III}]5007\AA\  over \halpha\ line ratio.  The black contours are the  I-band contours similar to  Fig. \ref{fig:halpha}. The white box denotes the borders of the slightly zoomed in \oiii\/\sii\ ionization map in panel a and the extinction map in Fig. \ref{fig:ebminv}.}
         \label{fig:ionization}
   \end{figure}

\subsection{Degree of ionization}\label{sec:ionization}
  \begin{figure*}[!ht]
   \centering
   \includegraphics[width=0.98\hsize]{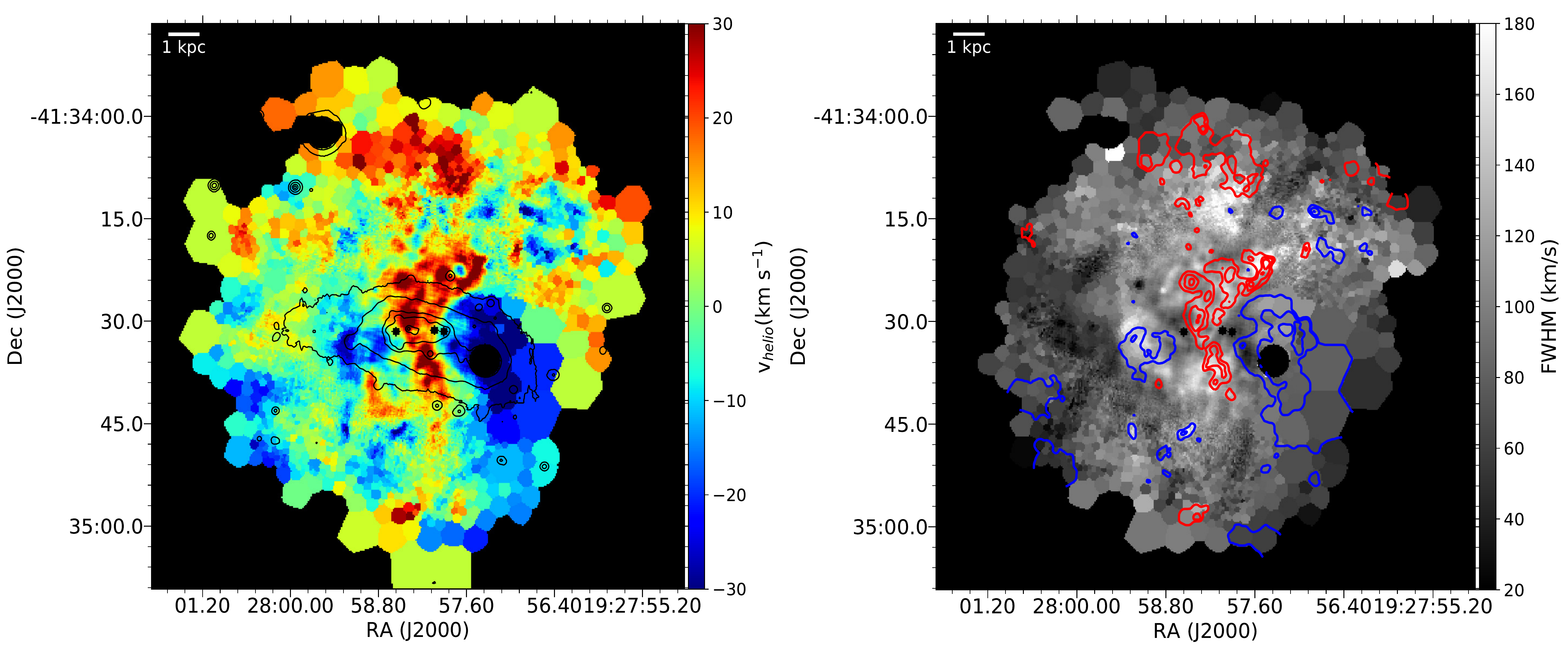}
      \caption{Kinematic information derived from a Gaussian fit to the \halpha\ emission with \emph{left} the \halpha\ velocity field.  The black contours are the contours of the I-band image reconstructed from the MUSE cube.  The \emph{right} figure shows the FWHM map of the \halpha\ line.  The instrumental FWHM is quadratically subtracted. Over-plotted in blue and red contours are the blue (-50, -30, -15 \kms) and red (15, 25, 35 \kms ) shifted velocity respectively. }
         \label{fig:kinematics}
   \end{figure*}

In \citet{Bik15} two ionization cones were identified based on the [\ion{S}{II}]/[\ion{O}{III}] line ratio. With the new dataset we are able to probe the physical properties even further out in the halo. Panel a in Fig \ref{fig:ionization} shows the deeper [\ion{O}{III}] over [\ion{S}{II}] map \citep[inverted with respect to][]{Bik15}.  Apart from the ionization cones to the north and the south and the very ionized central region, the more neutral areas east and west of the central star burst become more evident. This area in the halo coincides with the location of the older stellar populations, characterized by a strong \hbeta\ absorption (Sec. \ref{sec:hbeta}).

Another signature that becomes evident by tracing the outskirts of the halo is that the ionization increases again beyond 3.5 kpc, as shown in Fig. \ref{fig:ionization}, panel b, where the ratio of [\ion{O}{III}] and \halpha\ is plotted. These are the two strongest lines in our MUSE data, and therefore also trace the halo to the largest size.  In the area where both maps overlap, the same features are visible, suggesting that [\ion{O}{III}] over \halpha\ does mostly trace the ionization and to lesser extend  abundance variations in [O/H]. 

\citet{Pellegrini12} relate the \oiii/\sii\ ratio to the optical depth of LyC photons by applying photo ionization modelling. They show that at low optical depths the low-ionization line (\sii) is suppressed with respect to the high-ionization line. Only at higher optical depth, the  \oiii/\sii\  ratio will decrease towards the edges of the \ion{H}{ii} region, resembling a classical Str\"{o}mgren sphere and the \ion{H}{ii} region remains ionization bounded (LyC photons are captured).

In Fig. \ref{fig:ionization}, the central area of ESO 338 is highly ionized, showing very high ratios in both [\ion{O}{III}]/[\ion{S}{II}] and [\ion{O}{III}]/\halpha. Then the ratio drops to [\ion{O}{III}]/\halpha\ = 1.3 until 3.5 kpc from the center. At radial distances larger than 3.5 kpc, the ionization of the halo goes up again. This is especially visible in the north of the halo. The southern ionization cone also shows a very strong rise of the ionization after 3.5 kpc. In the northern halo, also an elongated feature of low-ionization is visible towards the north-west.

The decrease in [\ion{O}{III}] over \halpha\ at radii less than 3.5 kpc is analog to an HII region around a star, the O$^{++}$ zone is smaller than the H$^{+}$ due to the fact that few O$^{+}$ ionizing photons are available (h$\nu >$ 35.2 eV) than H ionizing photons (h$\nu >$ 13.6 eV). The radial increase at larger radii can be explained by several possibilities. The first explanation is that due to the fast dropping off gas density (Fig. \ref{fig:ne_sii}) the  mean-free path of the ionizing photons becomes large and the rate of recombinations does not balance anymore the rate of ionizations, resulting in a higher ionization fraction of the gas. This indicates that the halo is density bounded instead of ionization bound, and would suggest that ESO338 is leaking LyC photons  \citep[also found by][]{Leitet13}.  Also shocks can cause an increase in ionization and fast radiative shocks can result in emission of high ionization lines such as  \ion{Ne}{v} and \ion{He}{ii} \citep{Izotov01,Thuan05,Herenz17}. As will be discussed in Sect. \ref{sec:bpt}, we do find evidence for shocks around the central starburst , however, in the other regions where we find the increase in \oiii/\halpha\ ratio, the FWHM is rather low, suggesting that shocks might not be important here. A third possibility is presented by \citet{Binette09}, where they model gas turbulent mixing layers. Warm photoionized condensations are immersed in a hot supersonic wind resulting in turbulent dissipation and mixing, accelerating and heating of the gas.  This can lead to an increase of the \oiii/\halpha\ ratio. The condensations will be of higher density and therefore have a higher optical depth for the LyC photons, however, the lower density gas in which the condensations are immersed  still enables the escape of LyC photons.

Tracing even higher energetic photons are the emission lines from He$^{++}$ recombination (h$\nu >$ 54.4 eV). 
Over-plotted on panel a of Fig. \ref{fig:ionization} is a contour map of the \ion{He}{II} emission. Apart from the broad \ion{He}{II} emission peaking on some of the cluster positions (Sect \ref{sec:clusters}), there is a lot of spatially extended, diffuse \ion{He}{II} emission.  This diffuse emission has a narrow spectral profile and is of nebular origin.  We detect  diffuse \ion{He}{II} emission in the central area of the galaxy. In this area most of the young clusters, including the WR cluster (Sect \ref{sec:wrclusters}), are found. The \heii\ emission also overlaps with an area of high \oiii/\sii\ ratio, suggesting very high ionization. This indicates that the \heii\ is originating from photoionization by O stars and WR stars, that are also contributing to   the high  [\ion{O}{III}]/[\ion{S}{II}] ratio.

\subsection{Kinematics}\label{sec:kinematics}

In order to derive kinematic information such as the velocity and velocity dispersion, we apply a Voronoi pattern to the integrated \halpha\ emission line map with a minimum SNR of 20 and a maximum cell size of 900 $\square$  pixels (36 $\square$\arcsec). This pattern is applied to the reduced data cube and we fit the \halpha\ line in each Voronoi cell with a Gaussian profile. 

Fig. \ref{fig:kinematics} shows the results in both the velocity map as well as the map of the full-width at half maximum  (FWHM). The systemic velocity for ESO 338 is adopted to be 2841 \kms, as discussed in \citet{Bik15}.  The  FWHM of the instrument line spread function (LSF) at the observed wavelength of \halpha\ is calculated using the formulas given in Sect. 5.2 of \citet{Bacon17}. \citet{Bacon17} derive the FWHM of the LSF as function of wavelength  by fitting a large number  sky lines over the entire MUSE field of view. At wavelength of the redshifted \halpha\ emission (6625\AA), the FWHM  of the LSF is found to be 114.5 \kms. 
Because of the overall stability  of the instrument no significant changes in the spectral resolution are expected to occur. However, we checked this value on our data by measuring the FWHM of the Xe $\lambda$ 6595.56 \AA\  arc line (close to the  position of \halpha) in a reconstructed cube of an arc frame. The measured mean FWHM over the field of view of the instrument of the line is 114 \kms, with a standard deviation of 9 \kms, consistent with value derived by \citet{Bacon17}. We quadratically subtract the 114.5 \kms\ instrumental FWHM from the observed FWHM map. 

With the two datasets combined we can measure the velocity field and the FWHM as far as $\sim$40\arcsec (7 kpc) and both the velocity map and FWHM reveal many more structures than already discussed in  \citet{Bik15}.  In \citet{Bik15} we already presented the redshifted elongated features starting from the central area of the galaxy and extending north and south. \citet{Bik15} shows that the gas in the northern redshifted feature is highly ionized and conclude that these are outflows driven by feedback from the clusters in the central area of the galaxy. The new velocity maps shows that the north-western outflow, in contrast was what seen in \citet{Bik15}, does not extend to the end of the halo. It extends to a distance of 3.6 kpc. We will discuss the outflow feature in more detail in Sect. \ref{sec:wind}.

Further to the north-west of halo the velocity field turns chaotic with small patches showing blue shifted velocities and other patches showing redshifted velocities. The velocity difference between two patches can be as high as 60 \kms. North of the northern outflow feature, the velocity becomes again redshifted with a velocity of 30 \kms. Also in the southern part several blue and redshifted features can be seen in the velocity map. 

West of the starburst there is a large area with strongly blue shifted \halpha\ emission, with  velocities up to  -60 \kms. A similar blue shifted component is also visible in the velocity map of the \ion{H}{I} emission \citep{Cannon04}. This blue shifted region is not very bright in \halpha, while it is one of the brightest regions in \ion{H}{I}, rising the suggestion that this region is predominantly neutral. Inspecting the HST continuum images, we find that there is some star formation associated with this region, located just east of the  bright foreground star. This region, a bit offset from the main starburst, contains a handful of $10^4$ M$_{\sun}$ clusters between 1 and 6 Myr, making this region to stand out in the ionization maps (Fig. \ref{fig:ionization}) and  show nebular \heii\ emission. However, the star formation, and \heii\ emission, is very localized and at the edge of the blue shifted feature. The main part of this feature shows low ionization in the ionization maps (Fig. \ref{fig:ionization}). The feature is not very collimated and also the strong \ion{H}{I} emission suggests against it being an ionized outflow created by the stellar feedback.  We do not observe an increase in the FWHM. If this cloud would be in-falling in the galaxy, an increase in FWHM would be expected due to the super position of the in-falling cloud and the gas in the galaxy it self. Therefore we conclude that this cloud must be part of the galaxy ISM and possibly part of a perturbed disk of the galaxy.  We speculate that this cloud is a remnant of an in-falling cloud which triggered the current starburst. 

The  velocity field of ESO 338 is so perturbed that there is no clear signature of rotation in the halo.
Based on \halpha\ Fabry-Perot observations, \citet{Ostlin99,Ostlin01} derived rotation curves for a sample of BCG galaxies that included ESO 338.   They did not manage to derive a proper rotation curve for ESO 338 as the observed velocity gradient along the body of the galaxy is not symmetric due to the strong blue shifted emission. In our MUSE velocity map we have more information than in the Fabry-Perot maps, however, also in the MUSE maps, there is no real rotational profile detected.

The map of the velocity dispersion shows, similar to the velocity map the complexity of the ionized halo  of ESO 338. In general the FWHM shows rather large values in the halo ($\sim$ 100 \kms),  much higher than the thermal broadening. A detailed comparison between the FWHM map and the velocity map shows that the two redshifted feature interpreted as outflows also show high velocity dispersion, especially at the location where the redshifted emission end (see Sect. \ref{sec:wind}).
In the northern part of the halo, in between the red-shifted outflow and the redshifted regions further north, the FWM increases strongly. In the  northwestern part of the halo the chaotic velocity field also corresponds with a similar behaviour in FWHM map, where some patches have broad lines and others quite narrow.  More to the north, however, the velocities become ordered again, also resulting in a decrease of the FWHM.
Additionally, south and south east of the central area hosting all the young clusters an increase of the FWHM is observed in a ring like structure. 

In order to see whether ordered  or random motions dominate the kinematics and to compare ESO338 to high-redshift observations, we calculate the ratio $v_{\mathrm{shear}}$/$\sigma_0$. Where  $v_{\mathrm{shear}}$ is half the difference between the minimum and maximum velocity measuring the large scale motions in the galaxy and  $\sigma_0$ is the flux weighted average of the velocity dispersion, measuring the random motions in the galaxy.
We use the procedure described in \citet{Herenz16} to derive the two quantities.

We find that $v_{\mathrm{shear}}$ = 28.4 \kms, and $\sigma_0$ = 57.5 \kms, making the ratio $v_{\mathrm{shear}}$/ $\sigma_0$ = 0.5. These values suggest that the kinematics in ESO338 are highly dispersion dominated \citep{Glazebrook13}. This is a property found among more blue compact galaxies \citep{Ostlin01} as well as high-redshift star forming galaxies \citep{Newman13}. The energy released by the stellar feedback  increases the turbulence in these galaxies resulting in larger velocity dispersion.

\section{The ionized halo of ESO 338}\label{sec:halo}

The  \halpha\ line map (Fig. \ref{fig:halpha}) shows that the ionized gas around ESO 338 fills the entire FOV of the observations. In this section we study the halo in more detail and derive the total mass of the ionized gas as well as perform a spatially resolved BPT analysis of the ionized gas in the halo. Finally we look closer at the features identified as outflows in the halo of ESO 338.

\subsection{Ionized gas mass}\label{sec:HIImass}
  \begin{figure}
   \centering
   \includegraphics[width=\hsize]{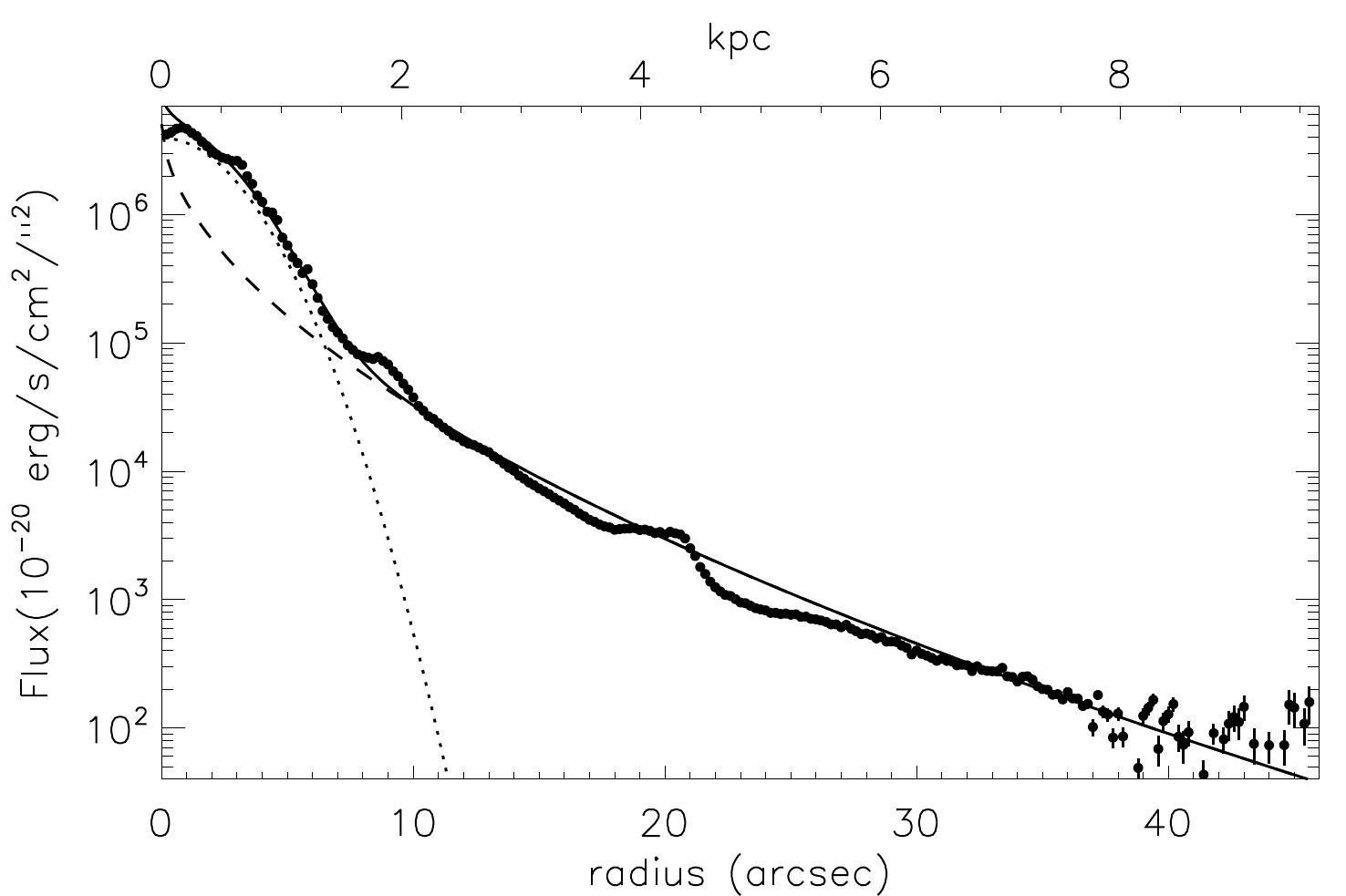}
      \caption{Radial surface brightness profile of the \halpha\ emission line. As central position, the position of cluster 23 has been chosen. Over plotted is a fit to the profile (solid line), consisting of a central Gaussian (dotted line) with $\sigma = 2.38 \arcsec (0.50$ kpc)  and a Sersic profile with an index of 1.8 to fit the extended halo with a scale length of 0.54\arcsec (0.11 kpc).} 
         \label{fig:sb_halpha}
   \end{figure}

In order to derive the mass of the ionized gas, we  construct a surface brightness (SB) profile of the \halpha\ emission (Fig. \ref{fig:sb_halpha}) by radially averaging the emission. The center of the profile is, as for the radial density profile, chosen to be the location of cluster 23. The SB profile is measured as far out as 9 kpc, with a dynamic range of about $10^6$. 

After a relatively constant central surface brightness, between 5 and 10\arcsec\ (2 kpc) the SB profile drops rapidly and  flattens beyond 10\arcsec\  and drops more monotonically to low values.  This SB profile can be fitted with a combination of a Gaussian profile in the center and a Sersic profile at larger radii. For the Gaussian in the central areas we measure a $\sigma$ of 2.38\arcsec\ (0.50 kpc). This emission likely is originating from the central starburst in ESO338. The outer halo can be well described by a Sersic profile with index n=1.8 and  a scale length of  0.54\arcsec (0.11) kpc. This profile describes the SB profile reasonably well down to faint SB levels (Fig. \ref{fig:sb_halpha}).  However, this is not a unique fit,  the profile can also be explained by a central Gaussian and two exponential profiles. An exponential profile with a short (2.2\arcsec) scale length, dominating between 10\arcsec\ and 20\arcsec\ and an exponential with a longer scale length (8\arcsec) representing the SB profile beyond 20\arcsec.
The derived surface brightness profile of ESO 338 is similar to what is found for other BCGs, where typically a central core with a exponential envelope is observed \citep[e.g.][\"Ostlin et al., in prep]{Papaderos02,Papaderos12}.

  \begin{figure*}[!t]
   \centering
   \includegraphics[width=\hsize]{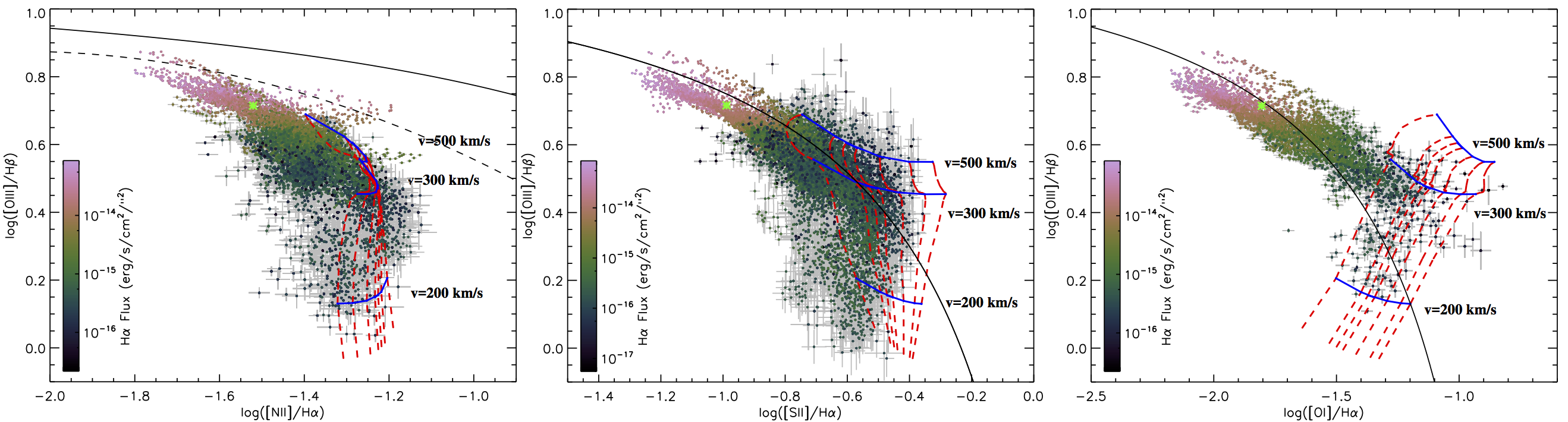}
      \caption{Spatially resolved BPT diagrams of the ionized gas around ESO 338. \emph{Left:} shows the "classic" BPT diagram [\ion{N}{II}]6583\AA/H$\alpha$ vs. [\ion{O}{III}]/H$\beta$ \citep{Baldwin81}, while the \emph{middle} panel shows the  [\ion{S}{II}]/H$\alpha$ vs [\ion{O}{III}]/H$\beta$ diagram, and the \emph{left} panel shows the   [\ion{O}{I}]/H$\alpha$ vs [\ion{O}{III}]/H$\beta$ \citep{Veilleux87}. The data points are colour-coded according to their \halpha\ flux in a logarithmic scaling. The areas where the \oiii\ or \halpha\ emission is saturated are excluded from the diagram.      The green asterisk  is the flux weighted average of all the data.The dashed line in the \nii\ diagram shows the \citet{Stasinska06} divisory line between star-forming galaxies and AGNs, while the solid line shows the divisory line from  \citet{Kewley01}. In the two other diagrams the solid lines separate the star forming galaxies from  AGN/LINERS according to \citet{Kewley01}. The red dashed (constant magnetic field) and blue solid (constant velocity) lines are predictions from fast radiative shock models by \citet{Allen08}, calculated for different magnetic field strength and shock velocity. Only the shock + precursor models with the SMC metallicity are plotted.}
         \label{fig:bpt}
   \end{figure*}

From the observed \halpha\  SB profile and the radial density profile, derived in Sect. \ref{sec:tempden}, we can derive the mass of the ionized halo around ESO 338. The mass of the ionized gas is related to the \halpha\ luminosity and electron  density as follows:

\begin{equation}
M_{HII} = \frac{\mu \ m_{H}\ L_{\mathrm{H}\alpha,0}}{h\ \nu_{\mathrm{H}\alpha}\ \alpha^{\mathrm{eff}}_{\mathrm{H}\alpha}\ n_e}
\label{eq:hm}
\end{equation}

where $\mu$ is the atomic weight, $m_H$ is the hydrogen mass,  $L_{\mathrm{H}\alpha,0}$ the extinction corrected \halpha\ luminosity, $h$  is Planck's constant, $\nu_{\mathrm{H}\alpha}$ is the frequency of
\halpha\, and $\alpha^{\mathrm{eff}}_{\mathrm{H}\alpha}$ is the Case B recombination coefficient for \halpha, and $n_e$ is the electron density.  For $\mu$, a values of 1.0 is chosen, and  $\alpha^{\mathrm{eff}}_{\mathrm{H}\alpha}$ is for a $1.2 \times10^4$ K gas equal to $9.835 \times10^{-14} \ \mathrm{cm}^3\  \mathrm{s}^{-1}$ \citep{Pequignot91}.

The \halpha\ luminosity is derived assuming a luminosity distance of 37.5 Mpc \citep{Ostlin98}.
The observed luminosity is corrected for extinction by making a radial E(B-V) profile assuming the \citet{Prevot84} extinction law. This profile is roughly constant through the halo with an average value of E(B-V) = 0.063 $\pm$ 0.05 mag. We derive this value by averaging the central 20\arcsec\ of the extinction radial profile. The error is chosen to reflect the radial profile further out and is derived by quadratically averaging the error between  0\arcsec\ and 40\arcsec.  

We do not measure the radial profile of the electron density as far out as we do for the \halpha\ SB profile. In order to extrapolate the electron density radial profile beyond 15\arcsec, we make an assumption about the shape of the radial profile.  Assuming the \halpha\ emission is emitted by recombination, the  \halpha\ luminosity is proportional to the square of the electron density ($L_{\mathrm{H}\alpha,0} \propto n_{e}^{2}$).  When we assume that the \halpha\ luminosity profile drops as a Sersic function with a certain scale length and index $n$= 1.8, the density radial profile would follow a Sersic function with the same index $n$, but with a scale length which is 3.48 times larger (1.88\arcsec). 
The absolute scaling of that Sersic profile was determined by fitting that function to the observed density profile between 6\arcsec\ and 11\arcsec\ (dashed line in Fig. \ref{fig:ne_sii}).

For  each radius we calculate Equation \ref{eq:hm}. We do the calculations only for the fit with the Sersic profile with n=1.8. For the 3 component profile with two exponentials, the derived values are very similar. Summing up over all radii gives a total hydrogen mass M(HII) = $3.0 \times 10^7$ \msun.  We also derive the total  \halpha\ luminosity corrected for extinction to be L$_{\mathrm{H}\alpha} = 3.56 \pm 0.06 \times 10^{41}$ erg s$^{-1}$. The error on the measured  L$_{\mathrm{H}\alpha}$ is dominated by the error on the extinction. The error on the total hydrogen mass is hard to determine as the derived value depends very strongly on the extrapolated radial profile for the electron density. The derived extinction corrected \halpha\ luminosity corresponds to a star formation rate (SFR) for ESO338 of 1.9 \msun\ yr$^{-1}$, adopting the calibration of \citet{Kennicutt12}. Note that this value strictly only valid under the assumption of a continuous star formation rate and no LyC leakage. The star formation rate derived from \halpha\ is sensitive to the stellar population with an age between 0 and 10 Myrs, with 3 Myrs as the the mean age of the  stellar population contributing to the \halpha\ emission \citep{Kennicutt12}. \citet{Ostlin03} and \citet{Adamo11} derive the star- and cluster formation history of ESO 338 and show that this is strongly rising with time.  However, at very young ages this becomes very uncertain as the SED fits to the youngest clusters are subject of many uncertainties in the models. In the case of a rising star formation history we would slightly over estimate the actual current  star formation rate. On the other hand effects such as absorption of LyC photons by dust, leakage of LyC photons will under estimate the derived star formation rate \citep[e.g.][]{Otifloranes10}.

We can compare the derived values to previous work by \citet{Ostlin99} and \citet{Ostlin09}. \citet{Ostlin99}  measured the \halpha\ luminosity from \halpha\ Fabry-Perot observations and found a  luminosity of $4.9 \times 10^{34}$ W ($4.9 \times 10^{41}$ erg s$^{-1}$). A similar value was found by  \citet{Ostlin09} from HST \halpha\ imaging. 
These values are  a bit higher than the total luminosity derived from the MUSE emission line map ($3.56 \pm 0.06 \times 10^{41}$ erg s$^{-1}$). However, the total mass we derive is five times lower than derived by \citet{Ostlin99}; $1.6\times 10^8$ \msun, assuming a constant density of $n_e = 10$ cm$^{-3}$. From the MUSE data we have derived a more realistic density profile, which is higher than 10 cm$^{-3}$ in the inner 12\arcsec\ of the halo. Inside this radius, also most of the \halpha\ flux is emitted. As the mass scales with the inverse of the electron density (Eq. \ref{eq:hm}), the derived mass will be lower with a higher electron density.

Also the SFR we derive is lower than the 3.2 \msun\ yr$^{-1}$  calculated by \citet{Ostlin01}. This is partly due to our lower \halpha\ luminosity and partly due to different conversions used. We use the conversion from \citet{Kennicutt12} assuming a  \citet{Kroupa03} IMF while \citet{Ostlin01} used their own conversion with a \citet{Salpeter55} IMF. 

Compared to the mass of the neutral hydrogen (1.4 $\pm$ 0.2) $\times 10^9$ \msun) derived from \ion{H}{I} observations by \citet{Cannon04}, the derive ionized gas mass only 1 \% of the total halo mass. This indicates that the halo around ESO 338 is predominantly neutral and could have implication for the escape of LyC photons from this galaxy.

  \begin{figure*}[!ht]
   \centering
   \includegraphics[width=0.97\hsize]{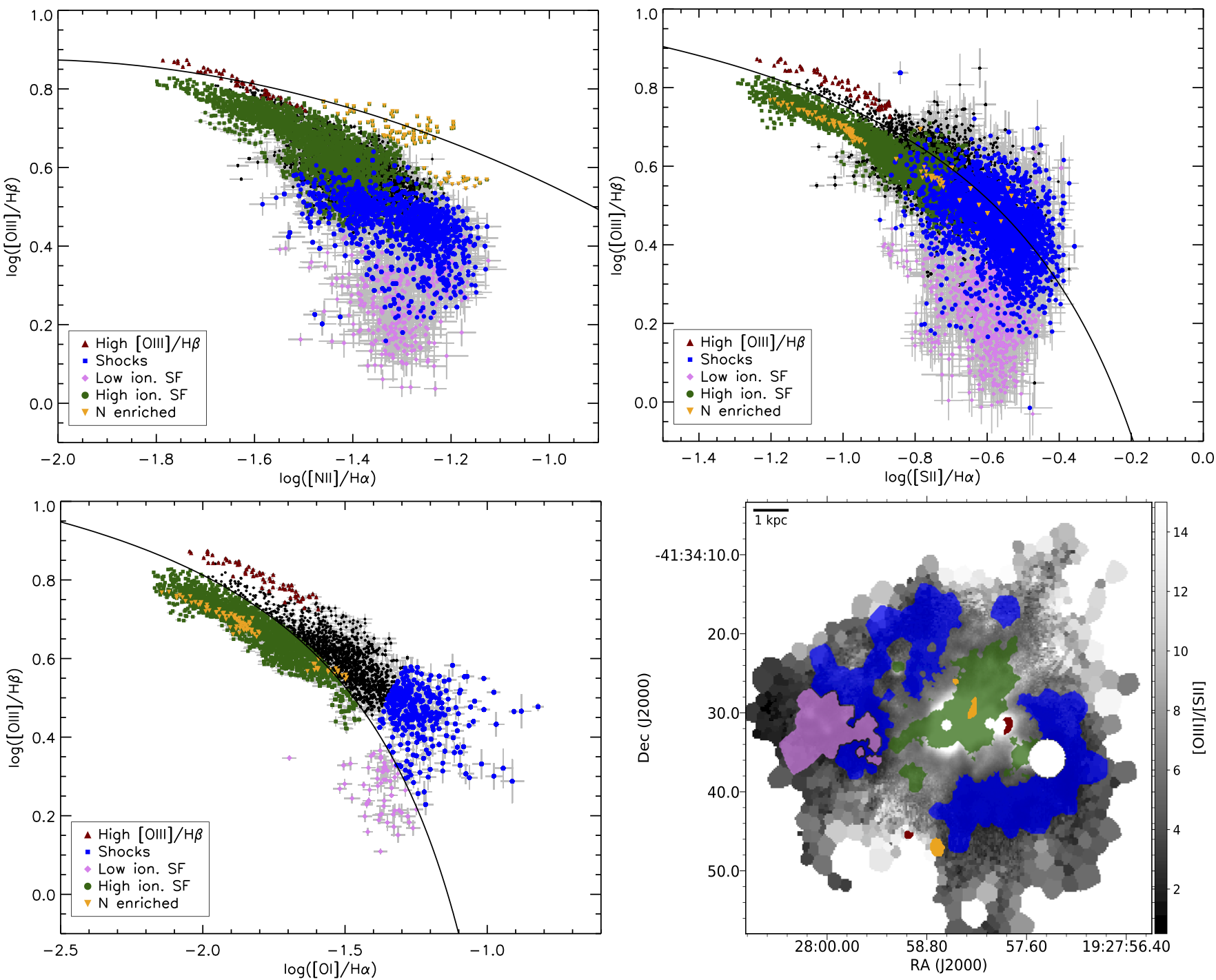}
      \caption{Spatially resolved BPT diagrams of ESO 338 (see Fig. \ref{fig:bpt}). The different colours represent different selections made in the BPT diagrams. {\bf Red:} Data points with a very high [\ion{O}{III}]/\hbeta\ ratio, selected in the SII diagram as being above the star formation line, but not originating in shocks log([\ion{S}{II}]/\halpha) < -0.8. {\bf Green:} High ionization (log([\ion{O}{III}]/\hbeta) > 0.4), star formation dominated gas (selected in the [\ion{O}{I}] diagram).  {\bf Blue:} Data points which cover the same area in the \oi\ diagram as the shock models of \citet{Allen08} and are located above the star formation line. {\bf Pink:} Low-ionization  (log([\ion{O}{III}]/\hbeta) < 0.4), star formation dominated gas selected in the \oi\ diagram. {\bf Orange:} Nitrogen enhanced gas detected as high [\ion{N}{II}]/\halpha\ ratio in the \nii\ diagram. {\bf Black:} Data points which do not fulfil any of the selection criteria. {\bf Bottom right:} In grayscale the [\ion{O}{III}]/[\ion{S}{II}] line ratio, tracing the ionization (Fig. \ref{fig:ionization}) with over-plottted the location of the different selections made in the BPT diagrams.}
         \label{fig:bpt_sel}
   \end{figure*}

\subsection{Spatially resolved BPT diagrams}\label{sec:bpt}

In order to gain  insight in the ionization processes of the ionized ISM and halo of ESO 338, we construct emission line ratio diagrams \citep[BPT diagrams,][]{Baldwin81,Veilleux87}. The location in the BPT diagram  is a function of many parameters, such as ionization, metallicity, electron density and hardness of the radiation field \citep[e.g.][]{Kewley06}, allowing us to trace these properties in a spatially resolved manner.

We construct three different BPT diagrams, all having the [\ion{O}{iii}]5700\AA/\hbeta\ ratio on the y-axis. We show the [\ion{N}{II}]6583\AA/\halpha\ (referred to as the [\ion{N}{II}] diagram), originally presented in  \citet{Baldwin81}.  Additionally, the BPT diagrams with [\ion{S}{II}](6713+6731)\AA/\halpha\ ([\ion{S}{II}] diagram) and  [\ion{O}{II}](6300)\AA/\halpha\ ([\ion{O}{I}] diagram) originally presented in \citet{Veilleux87} will be discussed.

In order to construct the different BPT diagrams, we apply  Voronoi binning to increase the signal-to-noise in the outskirts of the halo. For each diagram, we select the faintest emission line to calculate the Voronoi pattern (see Tab. \ref{tab:diag}). The Voronoi pattern is calculated ensuring a minimum signal-to-noise of 20 per cell and a maximum cell size of 900$\square$ pixel (36 $\square$\arcsec). The pattern is then applied to the other emission line maps required to construct the BPT diagram. This procedure results in a different Voronoi pattern for each BPT diagram as the faintest emission line is a different one for each of them.  For this reason the BPT diagram with [\ion{O}{I}] (the faintest line) has less data points than the other two diagrams. For the \sii\ diagram we have a total of 39049 cells, for the \nii\ diagram we have 26536, while for the \oi\ we have 21905 cells.

The line ratios calculated in each Voronoi cell represent a point in the BPT diagram (Fig. \ref{fig:bpt}). We also show in these diagrams the division lines between star forming galaxies and AGN/LINERs from \citet{Kewley01} and, only in the \nii\ diagram, also the divisory line from \citet{Stasinska06}.
As a green asterisk the location of the flux weighted average of all the data points is shown.  
The green asterisks show the position in the diagrams if the line ratios are calculated from summing all the measured flux in each emission line. 
The location of the green asterisk is consistent with previous determination of the position of ESO 338 in the BPT diagram \citep[][]{Bergvall85,Leitet13}. The green asterisks are all located in the upper left part of the BPT diagram, consistent with ESO 338 being a low-metallicity, high-ionization galaxy. 

In the [\ion{N}{II}] BPT diagram all data points are located well below the star formation line of \citet{Kewley06}, suggesting that photo ionization is the dominant ionization process.  However, looking at the [\ion{S}{II}] and [\ion{O}{I}] BPT diagrams, many points are located above the star formation line, highlighting the importance of other emission mechanisms than photoionization. One of those possible other mechanisms are shocks in the ISM.  Over-plotted in Fig. \ref{fig:bpt} are the model predictions for fast radiative shocks calculated with the MAPPINGS III code \citep{Allen08} for velocities between 200 and 500 km s$^{-1}$. We only show the precursor + shock models for  SMC metallicity, closest to the metallicity of ESO 338 \citep[12+log(O/H) $\approx$ 7.9,][]{Guseva12}.  These shock models  explain the location of the points with the highest \sii/\halpha\ and \oi/\halpha\ ratios  outside the star formation area in the BPT diagrams. \citet{Kewley06} show that increasing the electron density and adding a more extreme UV field (caused by e.g. WR stars or very hot O stars) increase both the [\ion{O}{III}]/\hbeta\ and [\ion{N}{II}]/\halpha\ ratio. A higher ionization parameter will, in low-metallicity  gas, result in an increase of the [\ion{O}{III}]/\hbeta\ ratio.

We have to be cautious when interpreting spatially resolved BPT diagrams, as the BPT diagram is originally used to separate star formation galaxies from AGN/LINER dominated galaxies using integrated emission line spectra.  \citet{Ercolano12} shows that, based on a 3D MHD simulation without shocks, a large fraction of the points may fall above the star formation line. The 2D projection of a 3D complex ISM can result in having both more neutral and highly ionized material along the line of sight. They show that this results in a high \oiii/\hbeta\ ratio and a high  \sii\ or \nii\ over \halpha\ ratio, moving the point to the upper right area of the BPT diagram mimicking the presence of shocks.

To study the nature of the different components in the BPT, we apply several selections in the different BPT diagrams and identify the spatial location of those points in the halo. Fig. \ref{fig:bpt_sel} shows the 3 BPT diagrams again, but now with different selections described below: 

\begin{itemize}
\item{\emph{High \oiii/\hbeta\ (red triangles):} In both the [\ion{S}{II}] and [\ion{O}{I}] diagrams we  find points with a very high \oiii/\hbeta\ and low \sii/\halpha\ or \oi/\halpha. In the [\ion{N}{II}] diagram they can be seen as a sequence above most of the data points between log([\ion{N}{II}]/\halpha) = -1.8 and -1.6.  We select these high [\ion{O}{III}]/\hbeta\  data points, located above the star formation locus.  We use the [\ion{S}{II}] diagram where they form the clearest sequence separated from the main cloud of points. The points are selected to have a  \oiii/\hbeta\ of at least 1.07 $\times$ the ratio of the separation between star formation and AGN. This ratio is chosen such that it separates the sequence from the other data points. Additionally, we select only those points which have  a  log [\ion{S}{II}]/\halpha\ ratio less than -0.87. 
}
\item{\emph{Shocks (blue squares):} We select all data points that overlap with the shock models of \citet{Allen08} and are located above the points  star formation line. We make this selection in the [\ion{O}{I}] diagram where the displacement from the star formation locus is the largest.}

\item{\emph{Low-ionization star formation (pink diamonds):} These points are selected in the \oi\ diagram with log \oiii/\hbeta\ < 0.4 and located within the star-formation locus. }

\item{\emph{High-ionization star formation (green dots):}   The points representing more highly ionized gas inside the star formation locus are selected in the  [\ion{O}{I}] diagram to have  [\ion{O}{iii}]/\hbeta\  > 0.4. 
}

\item{\emph{Nitrogen enhanced gas (orange upside down triangles):} This selection is made in the [\ion{N}{II}] diagram. From log([\ion{O}{III}]/\hbeta) ratios between  $\sim$0.6 and $\sim$0.75  several points with increased \nii/\halpha\ ratio compared to the bulk of the points at that \oiii/\hbeta\ ratio are present.  The high \oiii/\hbeta\ points are selected by selecting point following relations \oiii/\hbeta\ > -0.55\nii/\halpha\ -0.06 and \oiii/\hbeta\ > 0.65, while the low  \oiii/\hbeta\ points are selected  following  \oiii/\hbeta\ > -1.05\nii/\halpha\ -0.73 and \oiii/\hbeta\ < 0.6.
These points only stand out in the [\ion{N}{II}] diagram, in both the \sii\ and \oi\  diagrams they are located mostly in the star formation locus and clearly do not stand out as a separate sequence. This indicates that these points represent locations with enhanced nitrogen abundance (Sect. \ref{sec:nii}). }

\end{itemize}

As the selection is done in 3 different diagrams, points can be part of more than one selection. This concerns the high \oiii/\hbeta (red) and the nitrogen enriched (orange) data points. Al the other selections are done in the \oi\ diagram. The high \oiii/\hbeta\ data points do not overlap with any other selection, the offset between these points and the high-ionization star formation data points (green) is large enough to avoid overlap, also in the \oi\ diagram. The nitrogen enriched data points do overlap with other selections. Almost all of these points are classified as high-ionization star formation (green) in the \oi\ diagram. 

A large number of data points are not fulfilling any of the selection criteria and are plotted as black points in Fig \ref{fig:bpt_sel}. In the \oi\ diagram they are located above the division line, but do not overlap with the shock models of \citet{Allen08}. In the other two diagrams, the black points overlap with the green (high ionization star formation) and blue (shocks) points. These points trace star forming gas as well as  shocked gas or a mix of both. These points are likely located in between the central area dominated by star formation and the ring of shocked gas. 

In the bottom right panel of Fig. \ref{fig:bpt_sel} the location of the different selections discussed above are over-plotted on the gray-scaled plot of the  \oiii/\sii\ ionization map (Fig. \ref{fig:ionization}). It shows that the central area of the galaxy where all the YSCs are located is dominated by highly ionized gas, photo ionized by the massive stars in the clusters (green area). The green area extends towards the north-east and spatially overlaps with the outflow identified in the \halpha\ velocity field \citep{Bik15}, suggesting that the gas from the central areas is streaming out along the outflow cavity without producing strong shocks. The gas with the very high \oiii/\hbeta\ ratio (red) is mostly located near cluster 53.  This is one of the clusters identified as WR cluster in Sect. \ref{sec:wrclusters}. The \halpha\ emission towards the cluster is saturated (and masked out from the analysis), but just west of the cluster the gas shows the high \oiii/\hbeta. The WR stars in this cluster are likely responsible for this very high ionization.

The low-ionization gas in the star formation locus (pink) is located in the eastern part of the galaxy where a relatively large \hbeta\ absorption was found (Sect. \ref{sec:hbeta}). This  suggests a somewhat older stellar population emitting much less ionizing photons than the young population in the center. Also in the ionization maps (Fig. \ref{fig:ionization}), this area is characterised by line ratios indicative of low-ionization. 

The points outside the star formation locus and overlapping with the shock models (blue) are located in the outer regions of the halo.  The spatial distribution of the shocked gas is in the shape of a ring around the starburst, with two openings where the ionization cones (and northern outflow) are located \citep{Bik15}. This is suggestive of  the halo gas outside the starburst being shocked by the expanding HII regions created by the stellar feedback.  Note that following \citet{Ercolano12} the points above the star formation divisory line can also be explained by overlapping low-ionization and high ionization \HII\ regions.  The area we see the shocked gas also shows lower ionization. We do see, however, also an increase in FWHM, this would be more suggestive of the presence of shocks.

\subsubsection{Nitrogen enrichment}\label{sec:nii}

  \begin{figure}[!t]
   \centering
   \includegraphics[width=\hsize]{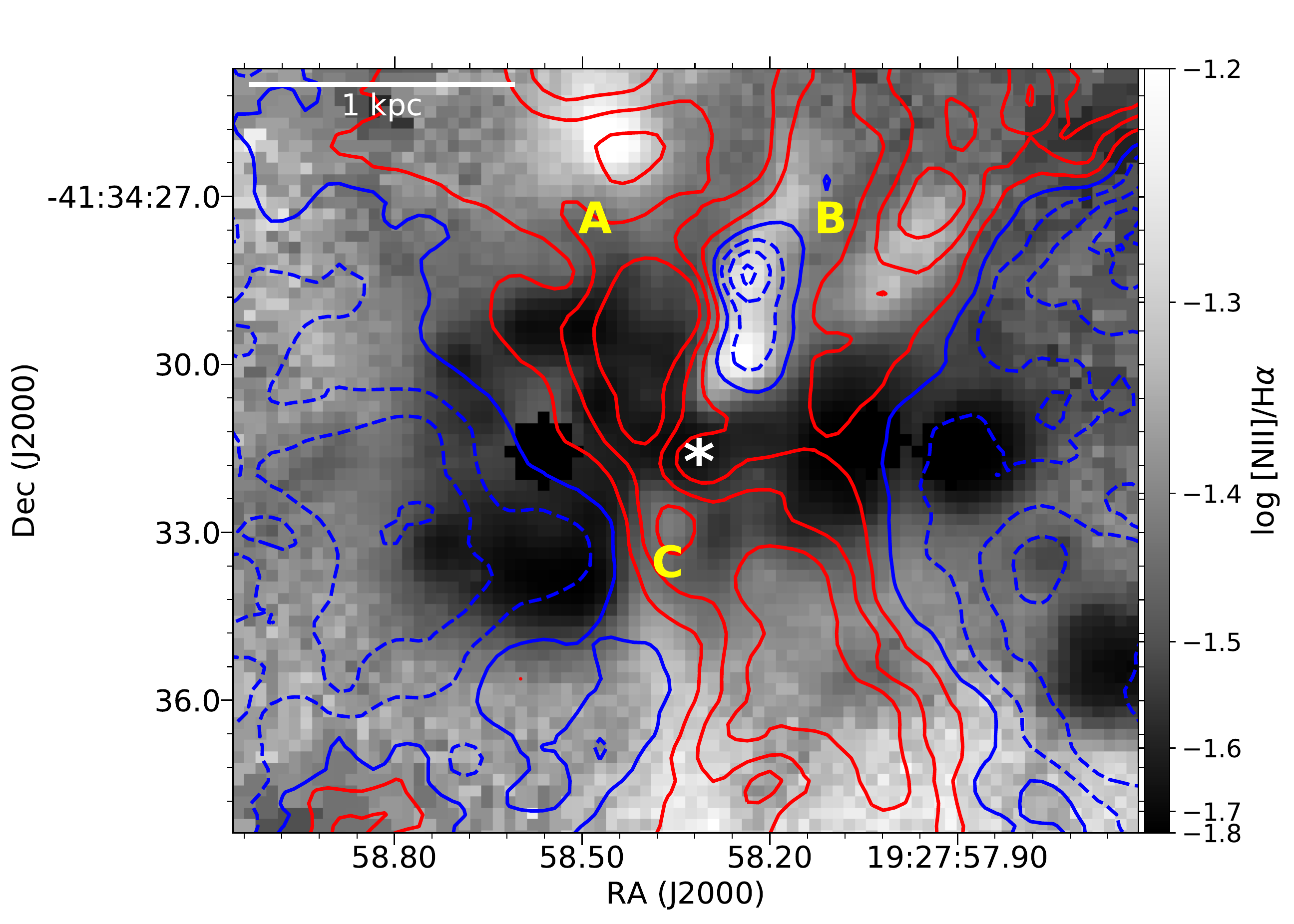}
      \caption{Map of the log(\nii/\halpha) ratio in the central areas of ESO 338, highlighting the features showing the increase in the \nii/\halpha\ ratio. 
 A v-shaped feature (B) as well as a circular blob (A) are seen north of the starburst clusters, while a fainter elongated feature (C) can be seen extending to the south. Overlayed in blue (-30, -20, -10, 0 \kms) and red (10, 20, 30 \kms) contours is the velocity field of the \nii 6583\AA\ line. 
 The eastern leg of feature B shows a strong blue shifted emission, which is not seen in \halpha.    The white asterisk is the possible location of the responsible cluster for the nitrogen enrichment.}
         \label{fig:nii}
   \end{figure}

To study the location of the gas with a high \nii/\halpha\ ratio in more detail, we show in Fig. \ref{fig:nii} the log(\nii (6583\AA)/\halpha) image of the central area of ESO 338. The dark areas in this figure show the low \nii/\halpha\ ratio at the location of where the gas is highly ionized (\ion{N}{ii} is ionized to \ion{N}{iii}). These areas correspond to the green data points in the BPT diagrams (high \oiii/\hbeta\ ratio) and corresponding to the location of the most massive star clusters.  Apart from the dark areas, we also can identify several bright areas with high \nii/\halpha\ ratio.
These \nii\ enriched areas  are marked with A, B and C in Fig. \ref{fig:nii}). 

Region A is north of most of the star clusters and is represented in the BPT diagram by the orange points around log(\oiii/\hbeta) $\sim$ 0.5 (Fig. \ref{fig:bpt_sel}). Region B has a v-shaped morphology and the brightest area of this region  corresponding to the orange points around  log(\oiii/\hbeta) $\sim$ 0.7 in the BPT diagram (Fig. \ref{fig:bpt_sel}). A much fainter enhancement (region C), elongated in southern direction is visible south of  region B.

Nitrogen enhancement has been observed in several BCG galaxies \citep[e.g.][]{James09,MonrealIbero10,MonrealIbero12,PerezMontero11}  towards WR clusters, suggesting that the emission is due to enrichment by WR stars \citep[or very massive stars above 100 \msun, which already show WR like stellar winds in a very early phase of their evolution, ][]{Smith16}. Region A has no corresponding continuum source detected in HST imaging, but is spatially coinciding with the outflow detected in \halpha.  The geometry of the two other features also suggests a relation to outflowing gas.  Also here, the emission is not directly co-located with a clusters, however, in between region B and C  several young clusters are detected.

To see if the enriched gas has the same velocity as the rest of the gas, we calculate the velocity map of \nii6583\AA, applying a Gaussian fit (identical to the \halpha\ line). The contours of the \nii\ velocity field are overlayed on the \nii/\halpha\ line map (Fig \ref{fig:nii}). Surprisingly, the eastern leg of region B shows a blue shifted velocity. This is not seen  in the \halpha\ velocity field (Fig. \ref{fig:kinematics}), where the gas is redshifted like the surrounding gas. We measure a maximum velocity of the \nii\ emission line  of -40 \kms (blue contours on Fig. \ref{fig:nii}), while the velocity measured from  \halpha\ is around +10 \kms.  We also find  a region of increased \nii6583\AA\  FWHM just east of region B, in between the blue shifted feature and the strongly redshifted feature just east of it, suggesting that we see both components overlapping resulting in a double line profile, not resolved by our MUSE spectroscopy.

The same features are also visible in the other \nii\ lines present in the MUSE spectrum. Both the 6549\AA\ and the much fainter auroral line 5755\AA\ show the same pattern as the 6583\AA\ line, both in their ratio to \halpha\ as well as in velocity, indicating that  this is not a data reduction artefact. 
We have inspected several line ratio maps (\oiii/\hbeta, \sii/\halpha, \oi/\halpha, \ion{He}{i}/\halpha, \ion{He}{ii}/\halpha, [\ion{Fe}{iii}]/\halpha) to see if enhancements in other elements other than nitrogen are present. None of the maps show the same features as the \nii/\halpha\ maps, suggesting that the gas is only enhanced in nitrogen. This makes the enhancement by supernovae unlikely, as supernova remnants are typically bright in [\ion{S}{II}] \citep{Blair12}.  The fact that these nitrogen enriched regions are highly ionized (high \oiii/\hbeta) supports the hypothesis that this is gas enriched by WR stars.

As the geometry of regions B and C are elongated, this gas could be outflowing gas. We can see in  Fig. \ref{fig:nii} that region B in the north and region C south are connected by an area with very low \nii/\halpha\ (high ionization). Inside this area of high ionization we find three bright clusters: clusters 27, 28 and 29 from the inner sample of \citet{Ostlin98}. All three clusters have masses of $\sim3 \times 10^5$ \msun.  Assuming a covering fraction of 0.5, the ages are 6 Myr for cluster 27 and 28 and 10 Myr for cluster 29.  These clusters are a bit too old for clusters containing WR stars at low metallicity.   We also do not see the  WR features towards these clusters. The clusters in which we identify the WR features (Sect. \ref{sec:wrclusters}) are not associated with the areas of N enrichment and they are also younger than the clusters we identify in between region B and C. An explanation for this apparent discrepancy can be related to the fact that the gas is outflowing. It will take some time before the gas is traveling away from  the clusters and by the time we observe the enriched gas, the WR stars have already disappeared.

\subsection{The galactic outflow}\label{sec:wind}

\begin{figure}[!t]
   \centering
   \includegraphics[width=1\hsize]{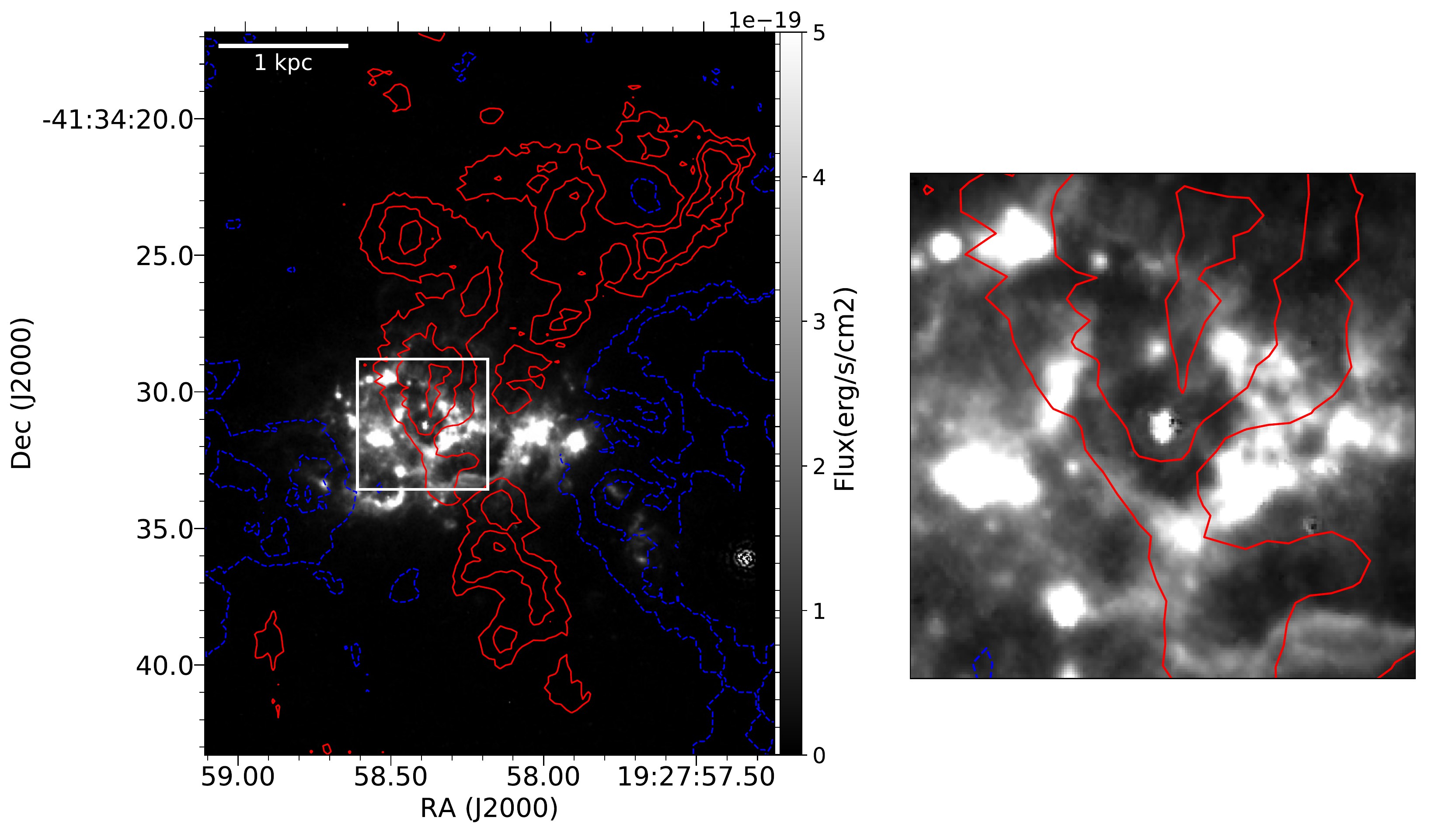}
         \caption{HST continuum corrected \halpha\ image taken from \citet{Ostlin09}, with the red shifted and blue shifted MUSE \halpha\ velocity map overlayed  as contours (Fig. \ref{fig:kinematics}). The red contours show the velocity contours of 20, 30 and 40 \kms, while the blue contours show  the velocities -10, -20 and-30 \kms. The zoom in shows the central 1 $\square$ kpc around cluster 23. The HST map shows the bubble around cluster 23 and the hole in that bubble towards the North.}  \label{fig:halpha_CL23}
   \end{figure}

We match the derived ISM properties from the MUSE data to the high-resolution HST data, in order to spatially characterize the outflow identified in the MUSE \halpha\ velocity map. Fig. \ref{fig:halpha_CL23}  shows the HST continuum subtracted \halpha\ image from \citet{Ostlin09} with the red shifted and blue shifted velocities from the MUSE \halpha\ velocity map overlayed as contours (Fig. \ref{fig:kinematics}).

At the high spatial resolution of the HST, the \halpha\ image shows the evacuated bubble around cluster 23. The location of cluster 23 is associated with the point source in the middle of the bubble.  The edges of the bubble do not show a uniform brightness and towards the north, the bubble seems to be open.
As already noted in \citet{Bik15}, the velocity contours in \halpha\ show that the northern outflow is a complex or probably multiple outflows. A comparison with the HST \halpha\ image suggests that cluster 23 is likely responsible for the most eastern part of the outflow, the base of that part of the outflow overlaps with the open side of the bubble (Fig. \ref{fig:halpha_CL23}). The stellar wind and supernova feedback form cluster 23 is responsible for creating the bubble. The fact that the bubble is open and a large scale outflow is driven from this bubble can be explained by the standard  model of galactic winds \citep[e.g.][]{Chevalier85,Heckman17}. Due to the mechanical energy input of the stellar winds and supernovae, a bubble is created and the gas inside the bubble is heated up and thermalised. Due to Rayleigh-Taylor instabilities, the bubble bursts and the hot gas streams out creating the galaxy scale outflow.

 \begin{figure*}[!ht]
   \centering
   \includegraphics[width=1\hsize]{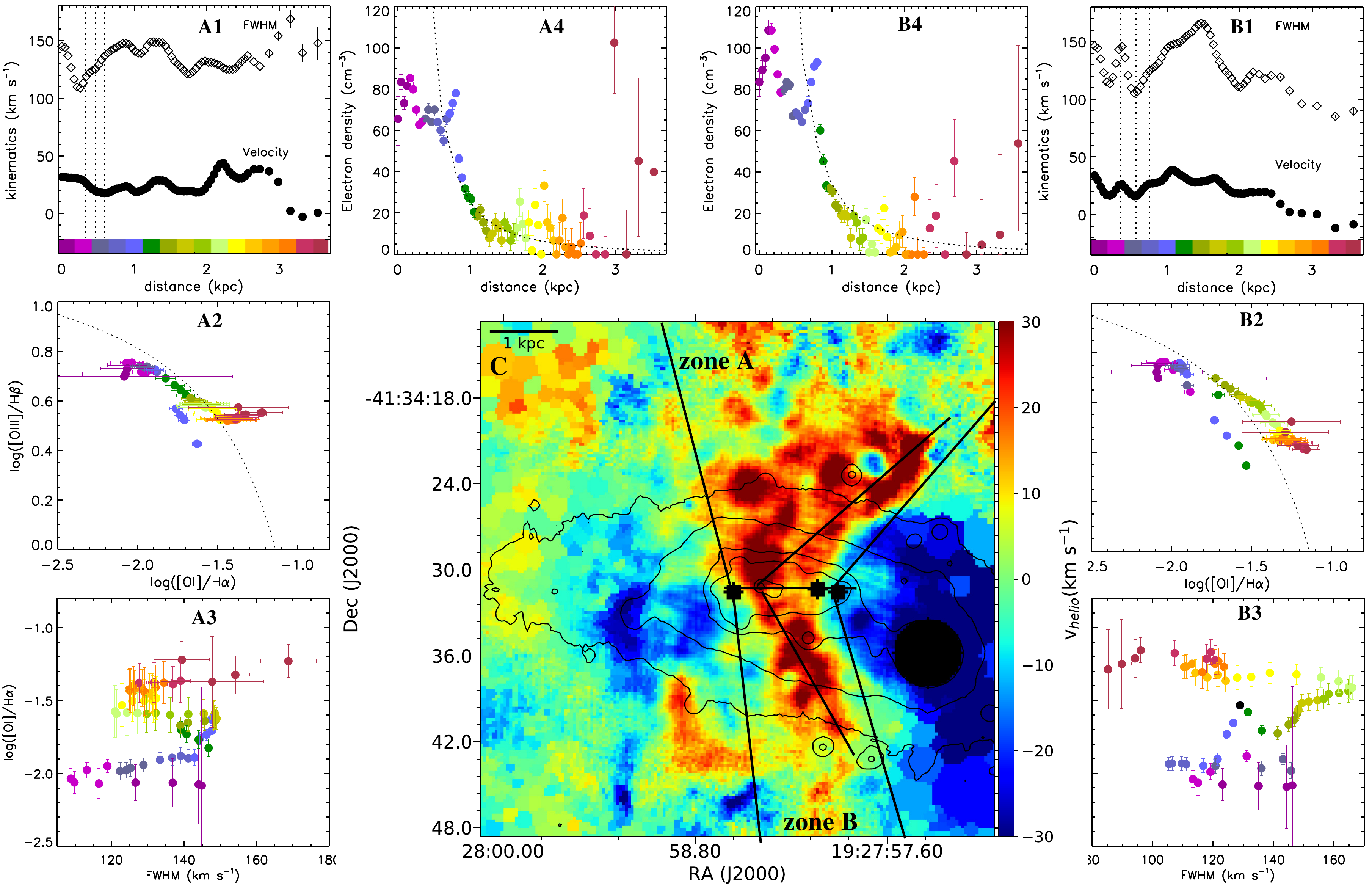}
         \caption{Derived outflow properties. \emph{Panel C} shows a zoomed-in \halpha\ velocity map, where we have outlined the northern (A) and southern (B) outflow zones in which we  derive the outflow properties. Additionally, two lines (with a length of 3.5 kpc each) are plotted from which we calculated an artificial long slit spectrum (5 pixels wide) in order to plot the kinematic information. The panels A1 to A4 show the extracted information for the northern outflow, and panels B1 to B4 the information for the southern outflow.  \emph{Panels A1 and B1} show the kinematic information (top FWHM, bottom velocity) extracted from the long slits as a function of distance from cluster 23. The colour bars at the bottom give the relation between colour and distance used in the remaining diagrams.  The vertical dotted lines mark the positions of the extracted FLAMES \halpha\ profiles (Fig. \ref{fig:flames_spec}). \emph{Panels A2 and B2} show the BPT diagram log[\ion{O}{i}]/\halpha\ vs log[\ion{O}{iii}]/\hbeta, with the ratios as a function of radius derived from the radial profiles of these lines.  \emph{Panels A3 and B3} plot the $\log$ [\ion{O}{i}]/\halpha\ ratio vs FWHM as a function of radius (colour-coded).  \emph{Panels A4 and B4} show the radial density profile derived from the \sii\ line ratio calculated using Pyneb. Over-plotted as dotted line is a $r^{-2}$ power-law as predicted for galactic scale outflow \citep{Chevalier85}.  
}\label{fig:outflow}

   \end{figure*}

To analyze the gas in the outflow cones in more detail, we derive radial profiles from the northern and southern outflow cone in several emission lines (as indicated in Fig. \ref{fig:outflow}c).  The radial profile is constructed starting from the position of cluster 23.  Our aim is to relate the measured velocities of the outflow with ISM properties such as the presence of shocks and the density profile. If we would derive the kinematic information from a similar radial profile as the ISM properties, the velocity variation would be heavily diluted at larger radii due to the high amount of structure in the velocity map. Therefore we create two pseudo long-slits with a width of 5 pixels (1\arcsec), one covering part of the northern outflow and one covering the southern outflow. This slit  has roughly the size of the seeing during the observations and preserves most of the details in the kinematics. The slits are indicated in Fig. \ref{fig:outflow}c. 

Panels A1 and B1 show the FWHM (open diamonds) and the velocity (filled circles) profile along those long-slits. The northern outflow (panel A1) shows an outflow velocity varying between 20 and 50 \kms\ over a projected distance of 3 kpc with the highest velocities measured between 2 and 3 kpc. The FWHM shows a peak near the center and then drops to a value of 100 \kms\ at 200 pc from cluster 23. This peak  is not spatially resolved in our MUSE data and likely originates from the gas inside or at the edge of the bubble around cluster 23, suggesting a very turbulent nature of the gas. The same can be seen in the southern outflow (panel B1).  At larger distances the FWHM in the northern outflow slowly increases again to values between 120 and 150 \kms. At a distance of 3 kpc, where the drop in velocity suggests that the outflow ends, the FWHM of the gas increases even further to values up to 170 \kms.  The southern outflow does not show this effect at the end of the outflow. In this outflow, after initial increase in velocity to $\sim$50 \kms\ at 1 kpc, the velocity slowly decreases and at 2.5 kpc really drops to $\sim$0 \kms. The FWHM varies strongly in the first 2 kpc, after which it slowly decreases to $\sim$80 \kms\ at 3 kpc distance from cluster 23.

 \begin{figure*}[!t]
   \centering
   \includegraphics[width=1\hsize]{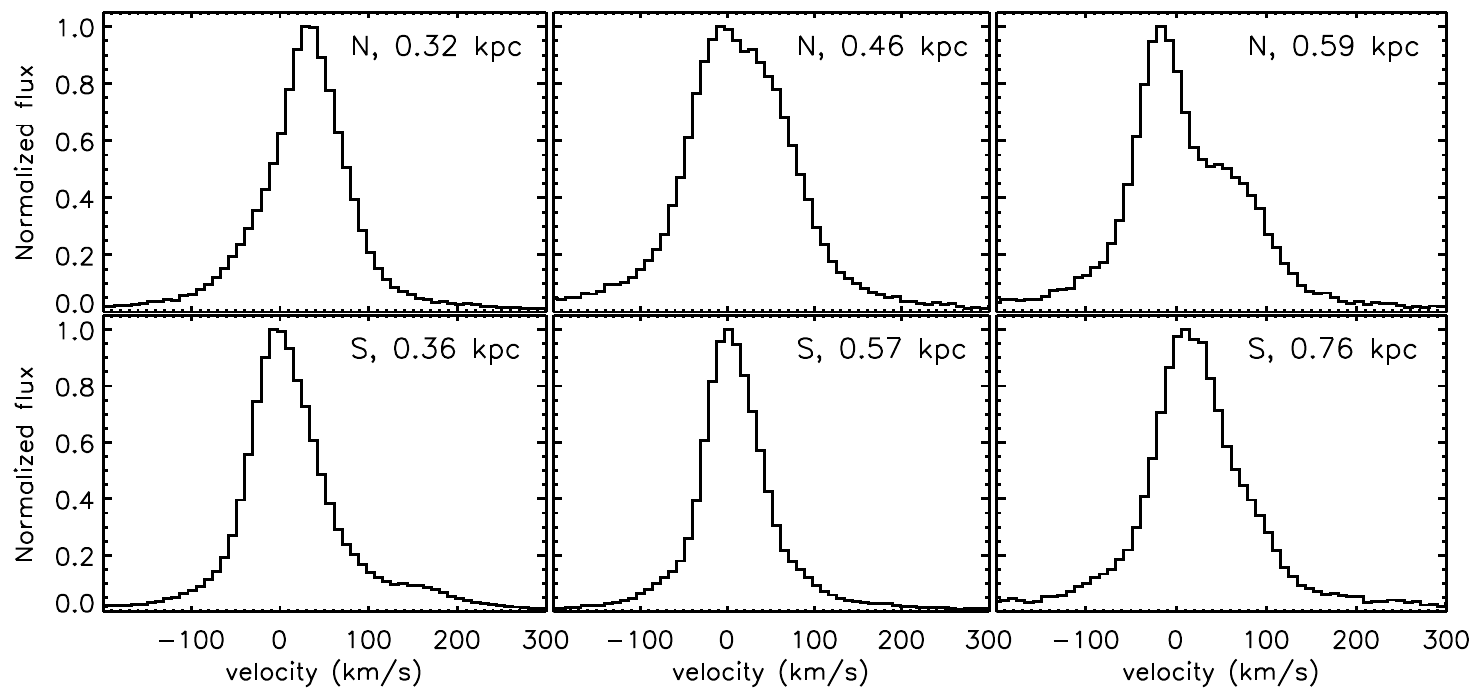}
         \caption{High resolution \halpha\ profiles of selected positions in the halo as observed with FLAMES. The top panels show the \halpha\ profile at 3 locations in the northern outflow, and the bottom panels show the corresponding profiles along the southern slit. These locations correspond to 3 points on the long slit defined in Fig. \ref{fig:outflow}. The distances are given along the slit and are measured from the position of cluster 23 as in Fig. \ref{fig:outflow}.}\label{fig:flames_spec}
   \end{figure*}

Panels A2 and B2 show the \oi\ BPT diagram (see also Fig \ref{fig:bpt}) of the radially averaged data in the outflow cones. The colours of the data points represent the distance from cluster 23 (see panel A1 and B1 for the relation between distance and colour).  Most points are located below the line separating star forming galaxies and AGNs \citep{Kewley01}. Close to cluster 23, the points are located in the upper left part of the diagram, suggesting high ionization.  The points move down with increasing radius following the separation line, after a certain distance the points cross the line and occupy the region of the BPT diagram dominated by shocked gas. 

For the northern outflow the points beyond 2.5 kpc are located in the shocked regime, which coincides with a strong increase in FWHM at the same distance. A similar effect can be seen in the southern outflow from 1.2 kpc, where at the largest FWHM, the points are located in the shocked regime of the BPT diagram.

These shocked areas are not seen in the spatially resolved BPT analysis (Fig. \ref{fig:bpt_sel}). Due to the radial averaging we are able to trace the gas properties out to much further distances than in the spatially resolved BPT analysis. The shocks appear outside the area covered in Fig. \ref{fig:bpt_sel}.

The relation between the log(\oi/\halpha) and the FWHM is shown in panels A3 and B3. These panels show indeed that when the \oi/\halpha\ ratio becomes larger, also the FWHM of the lines increases.  This is also seen for other galactic scale outflows \citep{Heckman90}. 
 This is especially clear in the northern outflow where the points at the end of the outflow show an increased FWHM and are also located above the separation line. In the souther outflow, the points between 1.2 and 2.5 kpc show a high FWHM, at these distances, the points are also above the separation line in the BPT diagram. Then at large radii, the FWHM slowly drops (as does the velocity), however, the  log(\oi/\halpha) remains high. 

Finally, figures A4 and B4 show the radial density profile derived from the \sii\ ratio for the two outflow cones. The dotted lines show the  the $r^{-2}$ relation expected for a galactic scale outflow \citep{Chevalier85,Heckman90}. The observed density profiles show similar behaviour as those observed by \citet{Heckman90}. In the center of ESO 338 they show a roughly flat profile, tracing the density of the gas inside the starburst region and the individual HII regions. The increase at very short distance from cluster 23 is tracing the ring around the bubble (see Fig \ref{fig:halpha_CL23}). After that the density sharply drops  consistently with the $r^{-2}$ relation observed in other starburst \citep{Heckman90} and with the theoretical production of galactic winds \citep{Chevalier85}.

In order to derive  the dynamical age of the outflows, we need to have an estimate of the inclination under which we see the galaxy and the outflow. As shown by the analysis of \citet{Ostlin99,Ostlin01}, the velocity field is so perturbed that, if even present, a rotation curve could not be derived, making it hard to get a handle on the inclination of the galaxy. \citet{Ostlin01} estimated the inclination of the disk of the galaxy to be 55$^{\circ}$. Assuming that the outflow is launched perpendicular to the disk plane, this would mean an outflow angle of 35$^{\circ}$ \citep[angle $\phi$ in][]{Heckman90}.

From our MUSE data we can derive some constraints; the observed velocities of the outflows are very low compared to other starburst galaxies The maximum velocity we measure is around 50 \kms, while for local starburst galaxies \halpha\ outflow velocities of several 100s of \kms\ are common \citep{Heckman90}. Assuming that outflow velocities are similar would suggest a low outflow angle. Additionally, we observe only red shifted velocities in the outflows. This excludes the possibility that the angle is close to 0. Typically the outflow is in the form of a cone with a finite opening angle \citep{Heckman90,Westmoquette09}. Both walls of the opening angle are showing red shifted velocity, suggesting an angle larger than 0.
The inclination derived by  \citet{Ostlin01} and the resulting 35$^{\circ}$ angle for the outflow is consistent with the above constraint. Therefore, for the remainder of the paper, we adopt an inclination of 55$^{\circ}$ used by \citet{Ostlin01} and the resulting  35$^{\circ}$ angle of the outflow under the assumption that the outflow is launched perpendicular to the disk.

For the northern outflow we measure a line of sight velocity between 30 and 50 \kms, this would then translate in a real velocity between 52 \kms\ and 87 \kms\ and spatial extend of 3.7 kpc. The line of sight velocity at the end of the outflow is 40 \kms, resulting in a real velocity of 70 \kms. Assuming a constant velocity with time, this results in a dynamical age of the outflow of $5 \times 10^7$ yr.  Comparing this dynamical age to the possible driving source cluster 23 shows that this cluster cannot be the driving sources as the age of cluster 23 is only $\sim$ 6 Myrs \citep{Ostlin07}. \citet{Ostlin03} and \citet{Adamo11} derive the  star and cluster formation history of the recent starburst and find the most recent burst started $\sim$ 40 Myr ago and showed an enhancement between 20 and 30 Myrs while the peak of the star formation activity is younger than 10 Myr.  These velocities would be consistent with the outflow being created at the beginning of the starburst in ESO 338 and the mechanical energy accelerating the ISM is now maintained by the new generation of star clusters.

However, we have to keep in mind that the measured centroid velocity  is derived from a Gaussian fit to a complex velocity profile, adding up all components present along the line of sight. Assuming that the \halpha\ emission mainly arrises from the walls of the outflow cones, the measured velocity is an average from both side of the outflow cone. Even if the cones move at relatively high speed, the average, centroid velocity can still be very low, especially, with the relatively low spectral resolution of MUSE where  low-flux high-velocity components will be easily missed. 

The central region of ESO 338 is also observed with the GIRAFFE/Argus integral field spectrograph of the FLAMES instrument mounted on the VLT.  The data, presented in \citet{Sandberg13}, centred at \halpha\  are obtained with a spectral resolution of R $\sim$ 12000, a factor of 5 higher than our MUSE spectra. We extract the FLAMES spectra at 3 location along the long slits defined in Fig. \ref{fig:outflow}.  The GIRAFFE/Argus field of view limits the analysis of the high resolution spectra to the central kpc of the galaxy. These spectra are presented in Fig. \ref{fig:flames_spec} and show a large diversity of line profiles. The northern outflows shows a double peaked profile with the red component increasing with velocity with distance. At a distance of 0.59 kpc from cluster 23, the two components have a projected velocity separation of 83 \kms, translating to a velocity separation of 143 \kms\ assuming an outflow angle of 35$^{\circ}$. The southern outflow shows a high velocity component at 0.36 kpc distance with a projected velocity of $\sim$150 \kms, which would result in 260 \kms\ corrected for outflow angle.  These high velocities observed in the FLAMES spectra also lead to much shorter dynamical time estimates. 

For example the high velocity component at a projected distance of 0.36 kpc from the center has a dynamical age of only 1.6 Myrs. This again illustrates the complexity of the outflow in ESO 338. The observed outflow complex is likely driven by many star clusters with different ages, leading to different components with a different dynamical age. This is especially true for the northern outflow, where also several spatially resolved components are visible on the velocity map confirms this picture. Both high spatial and spectral resolution observations would be needed to separate the different components and link them to individual clusters.

\section{Discussion}\label{sec:discussion}

In this paper we present deep MUSE integral field spectroscopy of the blue compact galaxy ESO 338-IG04. We find that the galaxy is surrounded by a large, low-density ionized halo extending as far as 9 kpc. We present a detailed analysis of the large ionized halo surrounding this galaxy and investigate the effect of the stellar cluster feedback  on the physical properties of the halo.

\subsection{The ionized ISM of ESO 338}\label{sec:ISM}

Using the optical emission lines we analyzed the ionized halo and ISM of ESO 338.  The central $\sim$1 kpc of the galaxy is dominated by the starburst and contains most of the massive young stellar clusters, including several with WR stars. The densities inside the region of the starburst are between 100 and 200 cm$^{-3}$.  Outside the central  $\sim$1 kpc  both the \halpha\ emission and the electron density drop fast. The \halpha\ surface brightness profile drops of with a Sersic index of 1.8. 

Analysis using the BPT diagram shows that the ionized gas becomes shocked outside the central starburst region.  Similar results have been found by \citet{Fensch16} in another dwarf galaxy.
We find the shocked gas to be located in ring around the starburst, with openings where we detect the outflows and ionization cones. We also see  an increase in FWHM of the \halpha\ line in the inner 3 kpc of the halo. The points overlap in the \oi\ BPT diagram with the fast shock models of \citet{Allen08}, more specifically those with shock velocities between 200 and 500 \kms. These high velocities are not measured in the  MUSE FWHM maps, however, analysis of the FLAMES spectra in some regions of the outflows shows the presence of faint, much higher velocity components. They are just not visible in the low resolution MUSE spectrum.  Additionally the shocks will be on small spatial scales, unresolved by our MUSE data. 

In our first paper on the MUSE data of ESO 338 we have identified two galactic scale outflows \citep{Bik15}.  Further analysis of the ISM lines in these outflows show a very complex spatial as well as velocity profile, suggesting that multiple clusters are responsible for driving the outflow. We also find evidence for shocks at the outer regions of the outflow. 

ESO 338 is a galaxy with similar properties to the samples discussed in \citep{Heckman15} where a detailed analysis of the outflow properties are done based on HST/COS absorption line spectroscopy. Several galaxies in their sample (e.g. NGC 5253, SBS 0335-052 and Haro 11) are classified as BCG, similarly as ESO 338.  The SFR (1.9 \msun\ yr$^{-1}$, Sect. \ref{sec:HIImass}) of ESO 338 is a  typical value measured in the sample of \citet{Heckman15}. The outflow properties, however, are more difficult to compare as \citet{Heckman15} uses absorption spectroscopy, where the centroid velocity of the warm ionized gas measured with the galaxy as background, while we measure the centroid of the emission line emitted in the same gas, but not restricted to the gas in front of the galaxy continuum. Following the trends between outflow velocity ($v_{\mathrm{out}}$ and SFR  an outflow velocity between $\sim$30 to $\sim$150 \kms could be expected for ESO 338. This is very similar to what we measure in the emission lines. 

Our emission line results can be better compared to \citet{Marlowe95} and \citet{Martin98} where similar star-forming galaxies are observed using \halpha\ imaging and long slit spectroscopy. Both studies derive similar dynamical ages as we do for the kpc scale  expanding super bubbles and filaments present in these dwarf galaxies.  They use the observed rotation curve to derive an estimate of the escape velocity and compare that to the observed outflow velocity in order to determine if the outflowing gas will escape the galaxy. They find that in most cases the outflowing gas has a velocity below the escape velocity, suggesting that outflowing gas will most likely not escape from the galaxy into the intergalactic medium (IGM).  It is therefore suggested that these galaxies have not yet developed full fledged galactic scale winds where gas is escaping in the IGM.

The super-bubbles are created by the combined mechanical output of the stellar winds and supernova, in which the energy input of the supernova thermalises the gas with a temperature up to  10$^7$ K. This gas will be very low density, but due to the high temperature still have a much larger pressure than the surrounding 10$^4$ K gas we trace with MUSE.  This hot super bubble will therefore expand in the warm gas and create shocks visible as gas with  an enhanced \oi/\halpha\ ratio similarly to the  ring of shocked gas seen in the ISM of ESO 338  at $\sim$ 3kpc from the center.
This ring can be caused by the expansion of the hot gas in a large super bubble surrounding the current starburst. 

For ESO 338 we do also observe what looks like a more developed galactic scale wind. However, the outflow velocities are still rather low, which could mean that also for ESO 338 this gas does not escape from the galaxy. Due to the disturbed velocity field of the halo of ESO 338 it is not possible to derive a rotation curve of this galaxy in order to trace the gravitational potential and derive an escape velocity \citep[see][]{Ostlin99}.  However, assuming a spherical gravitational potential we can derive an estimate of the escape velocity.

We use the relation for $v_\mathrm{esc}$ as function of radius and the mass of the galaxy derived by  \citet{Marlowe95}. Assuming a spherically symmetric isothermal potential with an outer cutoff at r$_\mathrm{max}$, the escape velocity at radius $r$ can be expressed as $v_\mathrm{esc}(r) = \sqrt 2 v_{circ} \times \sqrt{1+\ln( r_\mathrm{max} /r)}$. For $r_{max}$ we choose the Holmberg radius (V= 26.2 mag\arcsec$^2$ isophote) for ESO 338 derived by \citet{Bergvall02} to be 42\arcsec (7.6 kpc). The circular velocity is estimated using the virial theorem, again assuming a spherically symmetric potential, as $v_\mathrm{circ} = \sqrt{\frac{G\ M}{r_\mathrm{max}}}$.  For the mass we use the sum of the stellar mass \citep[$4\times 10^9$\msun,][]{Ostlin01} and the \ion{H}{I} mass \citep[$1.6\times 10^9$\msun,][]{Cannon04}. This results in $v_\mathrm{circ}$ = 55 \kms. The northern outflow has a de-projected length of 3.7 kpc. The escape velocity at this radius is $v_\mathrm{esc}$ = 102 \kms.  

This value is a bit higher than the de-projected outflow velocity measured for the northern outflow (87 \kms) and this would suggest that the outflowing gas does not escape.  However, the assumption of a spherically symmetric gravitational potential is not realistic and the real escape velocity could be different. On the other hand, no dark matter is included in the mass estimate, making the value a lower limit. 

From the radial velocity profile we can also see  that both the northern and southern outflow did not reach (yet) the edge of the ionized halo. This means that at this moment, no gas is being expelled from the galaxy, if the outflow develops towards the edge of the halo it could eventually. 
 
Because of the radial averaging we are able to trace the gas properties further out than in the spatially resolved analysis of the BPT diagram (Sect. \ref{sec:bpt}). The radial averaging enables us to identify shocked gas at the end of the outflow traced by both an increase in the velocity dispersion as well as an increase in the \oi/\halpha\ ratio. Analogous to the case of the expanding super-bubble, this can be explained by the hot outflowing gas interacting with the more quiescent halo gas  at the end of the outflow. Such an interaction would decelerate the outflow and create shocks at that location.  Another explanation for the FWHM increase at the end of the outflow is related to the fact that the outflow expands in a lower density medium. The outflowing gas becomes very turbulent.  Small scale shocks are created in the turbulent gas, resulting in an increase of the shock tracing lines.

\subsection{ WR stars and the origin of \ion{He}{II} emission}

We detect strong \heii\ emission in ESO 338. Towards the WR clusters we detect broad \heii\ 4686\AA\ emission, but we also find a spectroscopically narrow component. This narrow component is of more diffuse  origin. The presence of narrow \heii\ emission has been subject to multiple interpretations.  \heii\ emission requires the presence of a hard radiation field due to the high ionization potential of He$^{++}$ (54.4 eV).  

For galaxies at high redshift the narrow \heii\ emission is even seen as a tracer of pop III stars \citep{Cassata13} as it is much stronger than the fluxes expected for WR stars and gravitational cooling radiation produced by gas accretion. In local galaxies, narrow \heii\ emission is observed in the integrated spectra of several BCGs as well as low metallicity WR galaxies \citep[e.g.][]{Schaerer99,Kehrig08}, suggesting that the emission is being caused by ionization from WR stars.  \citet{Herenz17} on the other hand, find evidence that the \heii\ emission from the  super shell structure in SBS0335-52E  is  created by an enhanced UV field caused by shocks in the shell.  

In the case of ESO 338 we can see that the diffuse \heii\ emission is located inside the central starburst containing also the WR clusters. The contours of the \heii\ emission follow closely the areas of very high \oiii/\sii\ and \oiii/\halpha\ ratios, indication a high ionization \citep{Pellegrini12}.
 Due to the lower-density winds in low metallicity environments \citep[e.g.][]{Smith02}, not only WR stars are responsible for the \heii\ emission, but O stars could contribute significantly to the \heii\ ionizing photon flux.  
 It's spatial distribution makes it likely that the \heii\ emission in ESO 338 is caused by photoionization, with contributions from both the WR stars in the 4 clusters and massive O stars located in the younger clusters.

\subsection{The origin of the Nitrogen enhancement}

The analysis of the BPT diagram and the \nii/\halpha\ ratio revealed several areas with enhanced \nii\ emission (Sect. \ref{sec:nii}). We identify three distinct regions in the \nii/\halpha\ map (Fig. \ref{fig:nii}). Regions B and C  seem to be linked to two clusters right in the middle of these regions  with  ages slightly older than typical WR clusters. The other, most northern region (region A) is difficult to trace back to a cluster, but  has a velocity similar to the outflow detected in \halpha. 

In several BCGs nitrogen enhancements have been detected \citep[e.g.][]{James09,MonrealIbero10,MonrealIbero12,PerezMontero11,Kehrig13,Kumari18}, of which most studies suggest an association with WR stars. However, also other mechanisms are prosed to explain the observed enhancement in some galaxies.   \citet{PerezMontero11} find that the spatial extend of the nitrogen enhancement observed in several BCGs is too large to be explained by the enrichment due to WR winds, but the observed N/O values could be explained by the accretion of metal poor gas  following slow enrichment due star formation feedback \citep{Koppen05}. 

\citet{MonrealIbero10} show that in NGC5253 the ISM associated with one of the WR clusters shows a nitrogen enhancement. However not all the  WR clusters are found with an increase in nitrogen, \citet{MonrealIbero12} report a nitrogen enhancement   in NGC5253 without an associated WR cluster.   These observations made \citet{MonrealIbero12} come up with an evolutionary sequence, where the youngest clusters might not show nitrogen enhancement as the processed material needs to be transported to the stellar surface.  The second phase is when the WR stars expel the nitrogen enriched gas via winds and later as supernovae into the ISM around the cluster. The final phase is the mixing of the enriched material with the surrounding ISM  reaching a new chemical homogeneity.  This process is highly depend on the ISM properties and on the rate the gas is being diffused in the ISM \citep[e.g.][]{Williamson16} as well as the ISM properties themselves \citep{TenorioTagle96,deAvillez02}. Stellar evolution models show that rotation favours the appearance of processed material at the surface already for stars that are the main sequence \citep[e.g.][]{Kohler15}. Very massive stars (VMS) can also result in an enrichment of nitrogen at ages as early as 1 Myr.  \citet{Smith16} show that young stars significantly more massive than 100 \msun\ already exhibit very strong stellar wind, mimicking an appearance as a WR star,  making the picture more complicated than the proposed evolutionary sequence.

In contrast to the previous studies, where the nitrogen enhancement is mostly spatially associated with young cluster containing WR stars, in ESO 338 we find the emission displaced. We find  no clusters in the HST images at  location of the nitrogen enhancement  (Fig. \ref{fig:nii}). The elongated geometry of most of the emission suggests that it is outflowing gas expelled by the cluster(s) and currently flowing out with the rest of the outflowing gas. The difference measured in the velocity between the \nii\ and \halpha\ emission from one of the nitrogen enriched components remains intriguing. This enriched gas might be launched from a smaller sized super bubble where WR stars have enriched the gas inside the bubble. The different velocity suggest that not all the enriched gas is part of the main outflow complex but is launched in different direction. We do not see the enhancement in any other elements traceable with MUSE, the features are only seen in the three NII lines (5755\AA, 6583\AA\ and 6583\AA).  Moreover only these three lines show the blue shifted component. The emission lines of the other elements in the N-enhanced gas might be to faint to have enough contrast to see it against the bulk-motion of the un-enriched gas.  A high-resolution spectrum of the blue shifted component could reveal the other emission lines of this gas, allowing a better characterization of its properties, resulting in more clues on the origin of this gas.

The projected size of the longest leg of region B is 5.7\arcsec, corresponding to 1 kpc assuming a distance of 37.5 Mpc. By assuming the gas is launched perpendicular to the disk, an outflow angle of 35$^{\circ}$ and adopting the measured velocity of the blue shifted component (-40 \kms), we derive a dynamical age of $\sim10^7$ year. This is compatible with the age of the three clusters between region B and C. The three candidate clusters we detect in between region B and C  are between 6 and 10 Myrs old, supporting the scenario proposed by \citet{MonrealIbero12} that the appearance of the process material in the ISM takes some time and might even remain after the WR stars are gone, depending on the mixing timescales.

\subsection{Is ESO 338 a LyC leaker?}
The redshift of ESO 338 is too low to detect escaping LyC radiation directly via e.g. HST/COS as done for some Blue Compact Galaxies \citep[e.g.][]{Leitet11,Leitet13,Leitherer16,Puschnig17}. However, indirect evidence, based on residual emission under the usually saturated \ion{C}{ii} absorption line at 1036 \AA, suggests that ESO 338 is  a LyC leaker with an  escape fraction of $\approx$16 \% \citep{Leitet13}. The identified ionization channels in \citet{Bik15} support this indirect finding.  We observe an increase of the \oiii/\halpha\  and the \oiii/\sii\ ratios with distance from the center, indicating that the halo of ESO 338 is density bounded and likely optically thin for LyC photons.

The VLA 21 cm observations of \citet{Cannon04}  show that the ESO338 is embedded in a large halo of neutral hydrogen.  \citet{Cannon04} derive a total HI mass of  $(1.4 \pm 0.2) \times 10^9 $ \msun\ for the gas that is directly associated with the galaxy. The neutral gas is distributed roughly east-west, following the orientation of the projected semi-major axis in the optical. However, this study only detects the gas with column densities above $2\times 10^{20}$ cm$^{-2}$,  3 orders of magnitude above the column density where neutral hydrogen becomes opaque for LyC photons  \citep[N$_H$ = $1.6 \times10^{17}$ cm$^{-2}$,][]{Osterbrockbook}. The distribution of the lower-column density neutral gas is unknown.

In Sect. \ref{sec:HIImass} we derive a mass of the ionized gas based on the \halpha\ SB profile. The total ionized gas mass we find, $3.0 \times 10^7$ \msun\ is much lower than the neutral gas mass, suggesting that the halo around ESO 338 is predominantly neutral. This rises the question whether LyC photons are able to escape from this predominantly neutral halo. 

The observed spatial distribution of the ionized gas, however, is different from the distribution of the neutral gas. The ionized gas extends more to the north and south. The outflows are oriented roughly north and south and have transported a lot of ionized gas away from the body of the galaxy. 
Additionally, the ionization maps (Fig. \ref{fig:ionization}) suggest that  the ionized halo (as far as we can detect it) is not ionization bounded and LyC photons can escape.

Depending on the distribution of the lower-density neutral material, which is not detected in the \ion{H}{I} observations, there are two ways of letting the halo leak LyC photons. Firstly, if the low density neutral material shows the same geometric distribution as the gas detected in the \ion{H}{I} observations, the northern and southern parts of the halo serve as escape channels.  Secondly, if the lower-density neutral gas is more spherically distributed and the ionized halo is completely embedded in the neutral gas, a clumpy neutral medium, consisting of small dense clumps optically thick for LyC radiation  inside a much lower-density optically thin medium. The LyC radiation will then escape trough the low-density neutral medium, via ionization channels as seen in ESO 338 and other BCGs \citep{Zastrow11,Zastrow13,Herenz17}. The optically thick clumps will be the ones detected by the HI observations.

The ionization maps show that just east and west of the central starburst, the material is much more neutral than towards the north and the south (Fig. \ref{fig:ionization}b), and both the ionization cones and the outflows show a roughly bipolar geometry oriented north-south, with the main body of the galaxy oriented east-west.   This  suggests that the LyC photons preferentially leak via the north-south direction. This is similar to what is seen in e.g. M82 where the outflow is launched perpendicular to the disk of the galaxy \citep{Shopbell98}.

\section{Summary and conclusions}\label{sec:conclusions}

We present deep VLT/MUSE optical integral field observations of the blue compact galaxy ESO338-IG04. Based on analysis of the emission line spectra as well as HST observations of the cluster population we derive the following results:
\begin{enumerate}
\item{ESO 338 is surrounded by an extended ionized halo, which we can trace to 9 kpc distance from the center. We derive a total mass of the ionized gas of $3.0 \times 10^7$ \msun, by extrapolating the measured radial density profile as far out as the measured \halpha\ SB profile.}
\item{We identify 4 clusters showing the blue and/or red WR bump in their spectra. Re-fitting the observed SEDs of the clusters obtained from HST imaging data reveal that these clusters have masses above 10$^5$ \msun\ and ages between 3 and 6 Myrs, consistent with the observed WR features.}
\item{The ionization maps reveal the ionization channels detected earlier, but extending out even further to the north and the south to the edges of the observed halo. Additionally we find an increase with ionization when going to larger radii, suggesting that the halo of ESO 338 is density bounded. Diffuse \heii\ is found in the highly ionized central regions of the galaxy, consistent with ionization by WR and O stars.}
\item{The kinematical information derived from the \halpha\ line reveals a complex velocity structure of the halo. Similar to previous studies, no rotational motion could be identified in ESO 338. Instead the galaxy is a highly dispersion dominated galaxy with a $v_{\mathrm{shear}}/\sigma_{0}= 0.5$, similar to what is seen in high redshift star forming galaxies.}
\item{A spatially resolved analysis of the BPT diagram reveals that the gas in the central starburst as well as the inner $\sim$kpc of the outflows are dominated by photo-ionization. Towards one of the WR clusters we observe a very high \oiii/\hbeta\ ratio, indicative of the WR stars ionizing the surrounding gas to extreme levels. Outside the central regions, shocks start to become important. We find a ring of shocked gas around the central starburst. We interpret this as an expanding super bubble filled with hot gas created by the stellar wind and supernova feedback.}
\item{We find areas in the ISM with enriched nitrogen. This is consistent with the enrichment by WR stars. The regions with strong enrichment have an elongated spatial extend and show a velocity pattern different from that of the bulk of the gas detected in \halpha.  Towards the base of the features two stellar clusters are identified with ages slightly older than that of WR clusters. These clusters have formed their own small bubble with hot gas of which the nitrogen enriched gas is flowing out in a different direction than the bulk of the gas in the galaxy.}
\item{The \halpha\ kinematics reveals complex outflows towards the north and the south of the galaxy. Both outflows show a complex spatial as well as spectral structure, suggesting that multiple clusters are responsible for this. We find shocks at the end of the galactic outflows. This can be interpreted as hot outflowing gas colliding with the more quiescent warm ISM or by high density gas becoming kinematically unstable when entering low density gas. The turbulence created by these instabilities would then be responsible for the observed shocks.}
\item{The ISM and halo of ESO 338 is highly modified by the stellar feedback of the massive star clusters in ESO 338. All feedback mechanisms act  and become important in different areas and on different timescales. The mechanical feedback of the stars and supernovae is responsible for the shocks and outflows, modifying the spatial distribution of the ionized gas. Photo-ionization by the WR and O stars sets the ionziation structure of the halo and results in a density bounded halo, facilitating the escape of LyC photons.  Finally the ISM is chemically enriched by the WR winds.}
\end{enumerate}

\begin{acknowledgements}
 The anonymous referee is thanked for useful comments which helped to improve the quality of the paper. The authors thank Andreas Sandberg for sharing the reduced FLAMES data cubes.  We thank the ESO science verification team and the MUSE commissioning team for support and execution of the observations. G.\"O., M.H and A.A. acknowledge support of the  Swedish Research Council (Vetenskapsr\aa det)  and the Swedish National Space Board (SNSB). M.H. is Fellow of the Knut and Alice Wallenberg Foundation.

This research has made use of the NASA/IPAC Extragalactic Database (NED), which is operated by the Jet Propulsion Laboratory, California Institute of Technology, under contract with the National Aeronautics and Space Administration. This research has made use of NASA's Astrophysics Data System Bibliographic Services (ADS).
This research made use of Astropy, a community-developed core Python package for Astronomy \citep{astropy13}.
This research made use of APLpy, an open-source plotting package for Python \citep{Robitaille12}.
This research made use of MPFIT \citep{Markwardt09}.

    \end{acknowledgements}


\bibliographystyle{bibtex/aa}
\bibliography{MUSE338}

\begin{thebibliography}{121}
\expandafter\ifx\csname natexlab\endcsname\relax\def\natexlab#1{#1}\fi

\bibitem[{Adamo {et~al.}(2011)Adamo, {\"O}stlin, \& Zackrisson}]{Adamo11}
Adamo, A., {\"O}stlin, G., \& Zackrisson, E. 2011, \mnras, 417, 1904

\bibitem[{Adamo {et~al.}(2010{\natexlab{a}})Adamo, {\"O}stlin, Zackrisson,
  Hayes, Cumming, \& Micheva}]{Adamo10}
Adamo, A., {\"O}stlin, G., Zackrisson, E., {et~al.} 2010{\natexlab{a}}, \mnras,
  949

\bibitem[{Adamo {et~al.}(2017)Adamo, Ryon, Messa, Kim, Grasha, Cook, Calzetti,
  Lee, Whitmore, Elmegreen, Ubeda, Smith, Bright, Runnholm, Andrews, Fumagalli,
  Gouliermis, Kahre, Nair, Thilker, Walterbos, Wofford, Aloisi, Ashworth,
  Brown, Chandar, Christian, Cignoni, Clayton, Dale, de~Mink, Dobbs, Elmegreen,
  Evans, Gallagher, Grebel, Herrero, Hunter, Johnson, Kennicutt, Krumholz,
  Lennon, Levay, Martin, Nota, {\"O}stlin, Pellerin, Prieto, Regan, Sabbi,
  Sacchi, Schaerer, Schiminovich, Shabani, Tosi, Van~Dyk, \&
  Zackrisson}]{Adamo17}
Adamo, A., Ryon, J.~E., Messa, M., {et~al.} 2017, \apj, 841, 131

\bibitem[{Adamo {et~al.}(2012)Adamo, Smith, Gallagher, Bastian, Ryon,
  Westmoquette, Konstantopoulos, Zackrisson, Larsen, Silva-Villa, Charlton, \&
  Weisz}]{Adamo12}
Adamo, A., Smith, L.~J., Gallagher, J.~S., {et~al.} 2012, \mnras, 426, 1185

\bibitem[{Adamo {et~al.}(2010{\natexlab{b}})Adamo, Zackrisson, {\"O}stlin, \&
  Hayes}]{Adamo10SBS}
Adamo, A., Zackrisson, E., {\"O}stlin, G., \& Hayes, M. 2010{\natexlab{b}},
  \apj, 725, 1620

\bibitem[{Allen {et~al.}(2008)Allen, Groves, Dopita, Sutherland, \&
  Kewley}]{Allen08}
Allen, M.~G., Groves, B.~A., Dopita, M.~A., Sutherland, R.~S., \& Kewley, L.~J.
  2008, \apjs, 178, 20

\bibitem[{Astropy~Collaboration {et~al.}(2013)Astropy~Collaboration,
  Robitaille, Tollerud, Greenfield, Droettboom, Bray, Aldcroft, Davis,
  Ginsburg, Price-Whelan, Kerzendorf, Conley, Crighton, Barbary, Muna,
  Ferguson, Grollier, Parikh, Nair, Unther, Deil, Woillez, Conseil, Kramer,
  Turner, Singer, Fox, Weaver, Zabalza, Edwards, Azalee~Bostroem, Burke, Casey,
  Crawford, Dencheva, Ely, Jenness, Labrie, Lim, Pierfederici, Pontzen, Ptak,
  Refsdal, Servillat, \& Streicher}]{astropy13}
Astropy~Collaboration, T., Robitaille, T.~P., Tollerud, E.~J., {et~al.} 2013,
  \aap, 558, A33

\bibitem[{Bacon {et~al.}(2010)Bacon, Accardo, Adjali, Anwand, Bauer, Biswas,
  Blaizot, Boudon, Brau-Nogue, Brinchmann, Caillier, Capoani, Carollo, Contini,
  Couderc, Daguis{\'e}, Deiries, Delabre, Dreizler, Dubois, Dupieux, Dupuy,
  Emsellem, Fechner, Fleischmann, Fran{\c c}ois, Gallou, Gharsa, Glindemann,
  Gojak, Guiderdoni, Hansali, Hahn, Jarno, Kelz, Koehler, Kosmalski, Laurent,
  Le~Floch, Lilly, Lizon, Loupias, Manescau, Monstein, Nicklas, Olaya, Pares,
  Pasquini, P{\'e}contal-Rousset, Pell{\'o}, Petit, Popow, Reiss, Remillieux,
  Renault, Roth, Rupprecht, Serre, Schaye, Soucail, Steinmetz, Streicher,
  Stuik, Valentin, Vernet, Weilbacher, Wisotzki, \& Yerle}]{Bacon10}
Bacon, R., Accardo, M., Adjali, L., {et~al.} 2010, \procspie, 7735, 08

\bibitem[{Bacon {et~al.}(2017)Bacon, Conseil, Mary, Brinchmann, Shepherd,
  Akhlaghi, Weilbacher, Piqueras, Wisotzki, Lagattuta, Epinat, Guerou, Inami,
  Cantalupo, Courbot, Contini, Richard, Maseda, Bouwens, Bouch{\'e},
  Kollatschny, Schaye, Marino, Pello, Herenz, Guiderdoni, \& Carollo}]{Bacon17}
Bacon, R., Conseil, S., Mary, D., {et~al.} 2017, \aap, 608, A1

\bibitem[{Baldwin {et~al.}(1981)Baldwin, Phillips, \& Terlevich}]{Baldwin81}
Baldwin, J.~A., Phillips, M.~M., \& Terlevich, R. 1981, \pasp, 93, 5

\bibitem[{Bastian {et~al.}(2014)Bastian, Hollyhead, \&
  Cabrera-Ziri}]{Bastian14}
Bastian, N., Hollyhead, K., \& Cabrera-Ziri, I. 2014, \mnras, 445, 378

\bibitem[{Bergvall(1985)}]{Bergvall85}
Bergvall, N. 1985, \aap, 146, 269

\bibitem[{Bergvall \& {\"O}stlin(2002)}]{Bergvall02}
Bergvall, N. \& {\"O}stlin, G. 2002, \aap, 390, 891

\bibitem[{Bertin \& Arnouts(1996)}]{Bertin96}
Bertin, E. \& Arnouts, S. 1996, \aaps, 117, 393

\bibitem[{Bik {et~al.}(2015)Bik, {\"O}stlin, Hayes, Adamo, Melinder, \&
  Amram}]{Bik15}
Bik, A., {\"O}stlin, G., Hayes, M., {et~al.} 2015, \aap, 576, L13

\bibitem[{Binette {et~al.}(2009)Binette, Flores-Fajardo, Raga, Drissen, \&
  Morisset}]{Binette09}
Binette, L., Flores-Fajardo, N., Raga, A.~C., Drissen, L., \& Morisset, C.
  2009, \apj, 695, 552

\bibitem[{Blair {et~al.}(2012)Blair, Winkler, \& Long}]{Blair12}
Blair, W.~P., Winkler, P.~F., \& Long, K.~S. 2012, \apjs, 203, 8

\bibitem[{Bouwens {et~al.}(2015)Bouwens, Illingworth, Oesch, Caruana, Holwerda,
  Smit, \& Wilkins}]{Bouwens15}
Bouwens, R.~J., Illingworth, G.~D., Oesch, P.~A., {et~al.} 2015, \apj, 811, 140

\bibitem[{Brinchmann {et~al.}(2008)Brinchmann, Kunth, \& Durret}]{Brinchmann08}
Brinchmann, J., Kunth, D., \& Durret, F. 2008, \aap, 485, 657

\bibitem[{Calzetti {et~al.}(2015)Calzetti, Johnson, Adamo, Gallagher, Andrews,
  Smith, Clayton, Lee, Sabbi, Ubeda, Kim, Ryon, Thilker, Bright, Zackrisson,
  Kennicutt, de~Mink, Whitmore, Aloisi, Chandar, Cignoni, Cook, Dale,
  Elmegreen, Elmegreen, Evans, Fumagalli, Gouliermis, Grasha, Grebel, Krumholz,
  Walterbos, Wofford, Brown, Christian, Dobbs, Herrero, Kahre, Messa, Nair,
  Nota, {\"O}stlin, Pellerin, Sacchi, Schaerer, \& Tosi}]{Calzetti15}
Calzetti, D., Johnson, K.~E., Adamo, A., {et~al.} 2015, \apj, 811, 75

\bibitem[{Cannon {et~al.}(2004)Cannon, Skillman, Kunth, Leitherer, Mas-Hesse,
  {\"O}stlin, \& Petrosian}]{Cannon04}
Cannon, J.~M., Skillman, E.~D., Kunth, D., {et~al.} 2004, \apj, 608, 768

\bibitem[{Cappellari \& Copin(2003)}]{Cappellari03}
Cappellari, M. \& Copin, Y. 2003, \mnras, 342, 345

\bibitem[{Cassata {et~al.}(2013)Cassata, Le~F{\`e}vre, Charlot, Contini,
  Cucciati, Garilli, Zamorani, Adami, Bardelli, Le~Brun, Lemaux, Maccagni,
  Pollo, Pozzetti, Tresse, Vergani, Zanichelli, \& Zucca}]{Cassata13}
Cassata, P., Le~F{\`e}vre, O., Charlot, S., {et~al.} 2013, \aap, 556, A68

\bibitem[{Chevalier \& Clegg(1985)}]{Chevalier85}
Chevalier, R.~A. \& Clegg, A.~W. 1985, Nature, 317, 44

\bibitem[{Cooper {et~al.}(2008)Cooper, Bicknell, Sutherland, \&
  Bland-Hawthorn}]{Cooper08}
Cooper, J.~L., Bicknell, G.~V., Sutherland, R.~S., \& Bland-Hawthorn, J. 2008,
  \apj, 674, 157

\bibitem[{Crowther(2007)}]{Crowther07}
Crowther, P.~A. 2007, \araa, 45, 177

\bibitem[{de~Avillez \& Mac~Low(2002)}]{deAvillez02}
de~Avillez, M.~A. \& Mac~Low, M.-M. 2002, \apj, 581, 1047

\bibitem[{Diehl \& Statler(2006)}]{Diehl06}
Diehl, S. \& Statler, T.~S. 2006, \mnras, 368, 497

\bibitem[{Ercolano {et~al.}(2012)Ercolano, Dale, Gritschneder, \&
  Westmoquette}]{Ercolano12}
Ercolano, B., Dale, J.~E., Gritschneder, M., \& Westmoquette, M. 2012, \mnras,
  420, 141

\bibitem[{Fensch {et~al.}(2016)Fensch, Duc, Weilbacher, Boquien, \&
  Zackrisson}]{Fensch16}
Fensch, J., Duc, P.~A., Weilbacher, P.~M., Boquien, M., \& Zackrisson, E. 2016,
  \aap, 585, A79

\bibitem[{Ferland {et~al.}(2013)Ferland, Porter, van Hoof, Williams, Abel,
  Lykins, Shaw, Henney, \& {Stancil, P. C.}}]{Ferland13}
Ferland, G.~J., Porter, R.~L., van Hoof, P. A.~M., {et~al.} 2013, Revista
  Mexicana de Astronom{\'\i}a y Astrof{\'\i}sica Vol. 49, 49, 137

\bibitem[{Gil~de Paz {et~al.}(2003)Gil~de Paz, Madore, \&
  Pevunova}]{GildePaz03}
Gil~de Paz, A., Madore, B.~F., \& Pevunova, O. 2003, \apjs, 147, 29

\bibitem[{Glazebrook(2013)}]{Glazebrook13}
Glazebrook, K. 2013, \pasa, 30, 7

\bibitem[{Gonz{\'a}lez~Delgado \& Leitherer(1999)}]{GonzalezDelgadoI99}
Gonz{\'a}lez~Delgado, R.~M. \& Leitherer, C. 1999, \apjs, 125, 479

\bibitem[{Gonz{\'a}lez~Delgado {et~al.}(1999)Gonz{\'a}lez~Delgado, Leitherer,
  \& Heckman}]{GonzalezDelgadoII99}
Gonz{\'a}lez~Delgado, R.~M., Leitherer, C., \& Heckman, T.~M. 1999, \apjs, 125,
  489

\bibitem[{Guseva {et~al.}(2012)Guseva, Izotov, Fricke, \& Henkel}]{Guseva12}
Guseva, N.~G., Izotov, Y.~I., Fricke, K.~J., \& Henkel, C. 2012, \aap, 541,
  A115

\bibitem[{Guseva {et~al.}(2000)Guseva, Izotov, \& Thuan}]{Guseva00}
Guseva, N.~G., Izotov, Y.~I., \& Thuan, T.~X. 2000, \apj, 531, 776

\bibitem[{Heckman {et~al.}(2015)Heckman, Alexandroff, Borthakur, Overzier, \&
  Leitherer}]{Heckman15}
Heckman, T.~M., Alexandroff, R.~M., Borthakur, S., Overzier, R., \& Leitherer,
  C. 2015, \apj, 809, 147

\bibitem[{Heckman {et~al.}(1990)Heckman, Armus, \& Miley}]{Heckman90}
Heckman, T.~M., Armus, L., \& Miley, G.~K. 1990, \apjs, 74, 833

\bibitem[{Heckman \& Thompson(2017)}]{Heckman17}
Heckman, T.~M. \& Thompson, T.~A. 2017, arXiv:1701.09062

\bibitem[{Herenz {et~al.}(2016)Herenz, Gruyters, Orlitov{\'a}, Hayes,
  {\"O}stlin, Cannon, Roth, Bik, Pardy, Ot{\'\i}-Floranes, Mas-Hesse, Adamo,
  Atek, Duval, Guaita, Kunth, Laursen, Melinder, Puschnig, Rivera-Thorsen,
  Schaerer, \& Verhamme}]{Herenz16}
Herenz, E.~C., Gruyters, P., Orlitov{\'a}, I., {et~al.} 2016, \aap, 587, A78

\bibitem[{Herenz {et~al.}(2017)Herenz, Hayes, Papaderos, Cannon, Bik, Melinder,
  \& {\"O}stlin}]{Herenz17}
Herenz, E.~C., Hayes, M., Papaderos, P., {et~al.} 2017, \aap, 606, L11

\bibitem[{Izotov {et~al.}(2001)Izotov, Chaffee, \& Schaerer}]{Izotov01}
Izotov, Y.~I., Chaffee, F.~H., \& Schaerer, D. 2001, \aap, 378, L45

\bibitem[{Izotov \& Thuan(1999)}]{Izotov99}
Izotov, Y.~I. \& Thuan, T.~X. 1999, \apj, 511, 639

\bibitem[{James {et~al.}(2009)James, Tsamis, Barlow, Westmoquette, Walsh,
  Cuisinier, \& Exter}]{James09}
James, B.~L., Tsamis, Y.~G., Barlow, M.~J., {et~al.} 2009, \mnras, 398, 2

\bibitem[{Kehrig {et~al.}(2013)Kehrig, P{\'e}rez-Montero, Vilchez, Brinchmann,
  Kunth, Garc{\'\i}a-Benito, Crowther, Hern{\'a}ndez-Fern{\'a}ndez, Durret,
  Contini, Fern{\'a}ndez-Mart{\'\i}n, \& James}]{Kehrig13}
Kehrig, C., P{\'e}rez-Montero, E., Vilchez, J.~M., {et~al.} 2013, \mnras, 432,
  2731

\bibitem[{Kehrig {et~al.}(2018)Kehrig, Vilchez, Guerrero, Iglesias-P{\'a}ramo,
  Hunt, Duarte-Puertas, \& Ramos-Larios}]{Kehrig18}
Kehrig, C., Vilchez, J.~M., Guerrero, M.~A., {et~al.} 2018, \mnras, 480, 1081

\bibitem[{Kehrig {et~al.}(2015)Kehrig, Vilchez, P{\'e}rez-Montero,
  Iglesias-P{\'a}ramo, Brinchmann, Kunth, Durret, \& Bayo}]{Kehrig15}
Kehrig, C., Vilchez, J.~M., P{\'e}rez-Montero, E., {et~al.} 2015, \apjl, 801,
  L28

\bibitem[{Kehrig {et~al.}(2008)Kehrig, Vilchez, S{\'a}nchez, Telles,
  P{\'e}rez-Montero, \& Mart{\'\i}n-Gord{\'o}n}]{Kehrig08}
Kehrig, C., Vilchez, J.~M., S{\'a}nchez, S.~F., {et~al.} 2008, \aap, 477, 813

\bibitem[{Kennicutt \& Evans(2012)}]{Kennicutt12}
Kennicutt, R.~C. \& Evans, N.~J. 2012, \araa, 50, 531

\bibitem[{Kewley {et~al.}(2001)Kewley, Dopita, Sutherland, Heisler, \&
  Trevena}]{Kewley01}
Kewley, L.~J., Dopita, M.~A., Sutherland, R.~S., Heisler, C.~A., \& Trevena, J.
  2001, \apj, 556, 121

\bibitem[{Kewley {et~al.}(2006)Kewley, Groves, Kauffmann, \&
  Heckman}]{Kewley06}
Kewley, L.~J., Groves, B., Kauffmann, G., \& Heckman, T. 2006, \mnras, 372, 961

\bibitem[{Kohler {et~al.}(2015)Kohler, Langer, de~Koter, de~Mink, Crowther,
  Evans, Gr{\"a}fener, Sana, Sanyal, Schneider, \& Vink}]{Kohler15}
Kohler, K., Langer, N., de~Koter, A., {et~al.} 2015, \aap, 573, A71

\bibitem[{K{\"o}ppen \& Hensler(2005)}]{Koppen05}
K{\"o}ppen, J. \& Hensler, G. 2005, \aap, 434, 531

\bibitem[{Krause {et~al.}(2016)Krause, Charbonnel, Bastian, \&
  Diehl}]{Krause16}
Krause, M. G.~H., Charbonnel, C., Bastian, N., \& Diehl, R. 2016, \aap, 587,
  A53

\bibitem[{Kroupa(2001)}]{Kroupa01}
Kroupa, P. 2001, \mnras, 322, 231

\bibitem[{Kroupa \& Weidner(2003)}]{Kroupa03}
Kroupa, P. \& Weidner, C. 2003, \apj, 598, 1076

\bibitem[{Krumholz {et~al.}(2014)Krumholz, Bate, Arce, Dale, Gutermuth, Klein,
  Li, Nakamura, \& Zhang}]{Krumholz14PPVI}
Krumholz, M.~R., Bate, M.~R., Arce, H.~G., {et~al.} 2014, in Protostars and
  Planets VI

\bibitem[{Kumari {et~al.}(2018)Kumari, James, Irwin, Amor{\'\i}n, \&
  P{\'e}rez-Montero}]{Kumari18}
Kumari, N., James, B.~L., Irwin, M.~J., Amor{\'\i}n, R., \& P{\'e}rez-Montero,
  E. 2018, \mnras, 476, 3793

\bibitem[{Kunth {et~al.}(1988)Kunth, Maurogordato, \& Vigroux}]{Kunth88}
Kunth, D., Maurogordato, S., \& Vigroux, L. 1988, \aap, 204, 10

\bibitem[{Kunth \& {\"O}stlin(2000)}]{Kunth00}
Kunth, D. \& {\"O}stlin, G. 2000, Astronomy and Astrophysics Review, 10, 1

\bibitem[{Kunth \& Schild(1986)}]{Kunth86}
Kunth, D. \& Schild, H. 1986, Astronomy and Astrophysics (ISSN 0004-6361), 169,
  71

\bibitem[{Leitet {et~al.}(2013)Leitet, Bergvall, Hayes, Linn{\'e}, \&
  Zackrisson}]{Leitet13}
Leitet, E., Bergvall, N., Hayes, M., Linn{\'e}, S., \& Zackrisson, E. 2013,
  \aap, 553, A106

\bibitem[{Leitet {et~al.}(2011)Leitet, Bergvall, Piskunov, \&
  Andersson}]{Leitet11}
Leitet, E., Bergvall, N., Piskunov, N., \& Andersson, B.~G. 2011, \aap, 532,
  107

\bibitem[{Leitherer {et~al.}(2016)Leitherer, Hernandez, Lee, \&
  Oey}]{Leitherer16}
Leitherer, C., Hernandez, S., Lee, J.~C., \& Oey, M.~S. 2016, \apj, 823, 64

\bibitem[{Leitherer {et~al.}(1999)Leitherer, Schaerer, Goldader,
  Gonz{\'a}lez~Delgado, Robert, Kune, de~Mello, Devost, \&
  Heckman}]{Leitherer99}
Leitherer, C., Schaerer, D., Goldader, J.~D., {et~al.} 1999, \apjs, 123, 3

\bibitem[{Loose \& Thuan(1986)}]{Loose86}
Loose, H.~H. \& Thuan, T.~X. 1986, Star-forming dwarf galaxies and related
  objects, 73

\bibitem[{Luridiana {et~al.}(2015)Luridiana, Morisset, \& Shaw}]{Luridiana15}
Luridiana, V., Morisset, C., \& Shaw, R.~A. 2015, \aap, 573, 42

\bibitem[{Markwardt(2009)}]{Markwardt09}
Markwardt, C.~B. 2009, ASP Conf. Ser. 411, Astronomical Data Analysis Software
  and Systems XVIII (San Francisco: ASP), 411, 251

\bibitem[{Marlowe {et~al.}(1995)Marlowe, Heckman, Wyse, \&
  Schommer}]{Marlowe95}
Marlowe, A.~T., Heckman, T.~M., Wyse, R. F.~G., \& Schommer, R. 1995, \apj,
  438, 563

\bibitem[{Martin(1998)}]{Martin98}
Martin, C.~L. 1998, \apj, 506, 222

\bibitem[{McCall(2004)}]{McCall04}
McCall, M.~L. 2004, \aj, 128, 2144

\bibitem[{Meurer {et~al.}(1995)Meurer, Heckman, Leitherer, Kinney, Robert, \&
  Garnett}]{Meurer95}
Meurer, G.~R., Heckman, T.~M., Leitherer, C., {et~al.} 1995, \aj, 110, 2665

\bibitem[{Monreal-Ibero {et~al.}(2010)Monreal-Ibero, Vilchez, Walsh, \&
  Mu{\~n}oz-Tu{\~n}{\'o}n}]{MonrealIbero10}
Monreal-Ibero, A., Vilchez, J.~M., Walsh, J.~R., \& Mu{\~n}oz-Tu{\~n}{\'o}n, C.
  2010, \aap, 517, A27

\bibitem[{Monreal-Ibero {et~al.}(2012)Monreal-Ibero, Walsh, \&
  Vilchez}]{MonrealIbero12}
Monreal-Ibero, A., Walsh, J.~R., \& Vilchez, J.~M. 2012, \aap, 544, A60

\bibitem[{Newman {et~al.}(2013)Newman, Genzel, F{\"o}rster~Schreiber,
  Shapiro~Griffin, Mancini, Lilly, Renzini, Bouch{\'e}, Burkert, Buschkamp,
  Carollo, Cresci, Davies, Eisenhauer, Genel, Hicks, Kurk, Lutz, Naab, Peng,
  Sternberg, Tacconi, Wuyts, Zamorani, \& Vergani}]{Newman13}
Newman, S.~F., Genzel, R., F{\"o}rster~Schreiber, N.~M., {et~al.} 2013, \apj,
  767, 104

\bibitem[{Olofsson(1995)}]{Olofsson95}
Olofsson, K. 1995, \aaps, 111, 57

\bibitem[{Osterbrock \& Ferland(2006)}]{Osterbrockbook}
Osterbrock, D.~E. \& Ferland, G.~J. 2006, {Astrophysics of gaseous nebulae and
  active galactic nuclei} (Astrophysics of gaseous nebulae and active galactic
  nuclei, 2nd. ed. by D.E. Osterbrock and G.J. Ferland. Sausalito, CA:
  University Science Books, 2006)

\bibitem[{{\"O}stlin {et~al.}(2001){\"O}stlin, Amram, Bergvall, Masegosa,
  Boulesteix, \& M{\'a}rquez}]{Ostlin01}
{\"O}stlin, G., Amram, P., Bergvall, N., {et~al.} 2001, \aap, 374, 800

\bibitem[{{\"O}stlin {et~al.}(1999){\"O}stlin, Amram, Masegosa, Bergvall, \&
  Boulesteix}]{Ostlin99}
{\"O}stlin, G., Amram, P., Masegosa, J., Bergvall, N., \& Boulesteix, J. 1999,
  \aaps, 137, 419

\bibitem[{{\"O}stlin {et~al.}(1998){\"O}stlin, Bergvall, \&
  Roennback}]{Ostlin98}
{\"O}stlin, G., Bergvall, N., \& Roennback, J. 1998, \aap, 335, 85

\bibitem[{{\"O}stlin {et~al.}(2007){\"O}stlin, Cumming, \& Bergvall}]{Ostlin07}
{\"O}stlin, G., Cumming, R.~J., \& Bergvall, N. 2007, \aap, 461, 471

\bibitem[{{\"O}stlin {et~al.}(2009){\"O}stlin, Hayes, Kunth, Mas-Hesse,
  Leitherer, Petrosian, \& Atek}]{Ostlin09}
{\"O}stlin, G., Hayes, M., Kunth, D., {et~al.} 2009, \aj, 138, 923

\bibitem[{{\"O}stlin {et~al.}(2003){\"O}stlin, Zackrisson, Bergvall, \&
  R{\"o}nnback}]{Ostlin03}
{\"O}stlin, G., Zackrisson, E., Bergvall, N., \& R{\"o}nnback, J. 2003, \aap,
  408, 887

\bibitem[{Ot{\'\i}-Floranes \& Mas-Hesse(2010)}]{Otifloranes10}
Ot{\'\i}-Floranes, H. \& Mas-Hesse, J.~M. 2010, \aap, 511, A61

\bibitem[{Papaderos {et~al.}(2002)Papaderos, Izotov, Thuan, Noeske, Fricke,
  Guseva, \& Green}]{Papaderos02}
Papaderos, P., Izotov, Y.~I., Thuan, T.~X., {et~al.} 2002, \aap, 393, 461

\bibitem[{Papaderos {et~al.}(1996{\natexlab{a}})Papaderos, Loose, Fricke, \&
  Thuan}]{Papaderos96b}
Papaderos, P., Loose, H.~H., Fricke, K.~J., \& Thuan, T.~X. 1996{\natexlab{a}},
  \aap, 314, 59

\bibitem[{Papaderos {et~al.}(1996{\natexlab{b}})Papaderos, Loose, Thuan, \&
  Fricke}]{Papaderos96a}
Papaderos, P., Loose, H.~H., Thuan, T.~X., \& Fricke, K.~J. 1996{\natexlab{b}},
  \aaps, 120, 207

\bibitem[{Papaderos \& {\"O}stlin(2012)}]{Papaderos12}
Papaderos, P. \& {\"O}stlin, G. 2012, \aap, 537, A126

\bibitem[{Pasquali {et~al.}(2011)Pasquali, Bik, Zibetti, Ageorges, Seifert,
  Brandner, Rix, J{\"u}tte, Knierim, Buschkamp, Feiz, Gemperlein, Germeroth,
  Hofmann, Laun, Lederer, Lehmitz, Lenzen, Mall, Mandel, M{\"u}ller, Naranjo,
  Polsterer, Quirrenbach, Sch{\"a}ffner, Storz, \& Weiser}]{Pasquali11}
Pasquali, A., Bik, A., Zibetti, S., {et~al.} 2011, \aj, 141, 132

\bibitem[{Pellegrini {et~al.}(2012)Pellegrini, Oey, Winkler, Points, Smith,
  Jaskot, \& Zastrow}]{Pellegrini12}
Pellegrini, E.~W., Oey, M.~S., Winkler, P.~F., {et~al.} 2012, \apj, 755, 40

\bibitem[{Pequignot {et~al.}(1991)Pequignot, Petitjean, \&
  Boisson}]{Pequignot91}
Pequignot, D., Petitjean, P., \& Boisson, C. 1991, \aap, 251, 680

\bibitem[{P{\'e}rez-Montero {et~al.}(2011)P{\'e}rez-Montero, Vilchez,
  Cedr{\'e}s, H{\"a}gele, Moll{\'a}, Kehrig, D{\'\i}az, Garc{\'\i}a-Benito, \&
  Mart{\'\i}n-Gord{\'o}n}]{PerezMontero11}
P{\'e}rez-Montero, E., Vilchez, J.~M., Cedr{\'e}s, B., {et~al.} 2011, \aap,
  532, A141

\bibitem[{Portegies~Zwart {et~al.}(2010)Portegies~Zwart, McMillan, \&
  Gieles}]{Portegies10}
Portegies~Zwart, S.~F., McMillan, S. L.~W., \& Gieles, M. 2010, \araa, 48, 431

\bibitem[{Prevot {et~al.}(1984)Prevot, Lequeux, Prevot, Maurice, \&
  Rocca-Volmerange}]{Prevot84}
Prevot, M.~L., Lequeux, J., Prevot, L., Maurice, E., \& Rocca-Volmerange, B.
  1984, \aap, 132, 389

\bibitem[{Puschnig {et~al.}(2017)Puschnig, Hayes, {\"O}stlin, Rivera-Thorsen,
  Melinder, Cannon, Menacho, Zackrisson, Bergvall, \& Leitet}]{Puschnig17}
Puschnig, J., Hayes, M., {\"O}stlin, G., {et~al.} 2017, \mnras, 469, 3252

\bibitem[{Robertson {et~al.}(2015)Robertson, Ellis, Furlanetto, \&
  Dunlop}]{Robertson15}
Robertson, B.~E., Ellis, R.~S., Furlanetto, S.~R., \& Dunlop, J.~S. 2015,
  \apjl, 802, L19

\bibitem[{Robitaille \& Bressert(2012)}]{Robitaille12}
Robitaille, T. \& Bressert, E. 2012, Astrophysics Source Code Library

\bibitem[{Salpeter(1955)}]{Salpeter55}
Salpeter, E.~E. 1955, \apj, 121, 161

\bibitem[{Sandberg {et~al.}(2013)Sandberg, {\"O}stlin, Hayes, Fathi, Schaerer,
  Mas-Hesse, \& Rivera-Thorsen}]{Sandberg13}
Sandberg, A., {\"O}stlin, G., Hayes, M., {et~al.} 2013, \aap, 552, 95

\bibitem[{Schaerer {et~al.}(1999)Schaerer, Contini, \& Kunth}]{Schaerer99}
Schaerer, D., Contini, T., \& Kunth, D. 1999, \aap, 341, 399

\bibitem[{Schlafly \& Finkbeiner(2011)}]{Schlafly11}
Schlafly, E.~F. \& Finkbeiner, D.~P. 2011, \apj, 737, 103

\bibitem[{Searle \& Sargent(1972)}]{Searle72}
Searle, L. \& Sargent, W. L.~W. 1972, \apj, 173, 25

\bibitem[{Shopbell \& Bland-Hawthorn(1998)}]{Shopbell98}
Shopbell, P.~L. \& Bland-Hawthorn, J. 1998, \apj, 493, 129

\bibitem[{Smith {et~al.}(2016)Smith, Crowther, Calzetti, \& Sidoli}]{Smith16}
Smith, L.~J., Crowther, P.~A., Calzetti, D., \& Sidoli, F. 2016, \apj, 823, 38

\bibitem[{Smith {et~al.}(2002)Smith, Norris, \& Crowther}]{Smith02}
Smith, L.~J., Norris, R. P.~F., \& Crowther, P.~A. 2002, \mnras, 337, 1309

\bibitem[{Stasinska {et~al.}(2006)Stasinska, Cid~Fernandes, Mateus, Sodr{\'e},
  \& Asari}]{Stasinska06}
Stasinska, G., Cid~Fernandes, R., Mateus, A., Sodr{\'e}, L., \& Asari, N.~V.
  2006, \mnras, 371, 972

\bibitem[{Strickland \& Heckman(2009)}]{Strickland09}
Strickland, D.~K. \& Heckman, T.~M. 2009, \apj, 697, 2030

\bibitem[{Tenorio-Tagle(1996)}]{TenorioTagle96}
Tenorio-Tagle, G. 1996, \aj, 111, 1641

\bibitem[{Thompson {et~al.}(2016)Thompson, Quataert, Zhang, \&
  Weinberg}]{Thompson16}
Thompson, T.~A., Quataert, E., Zhang, D., \& Weinberg, D.~H. 2016, \mnras, 455,
  1830

\bibitem[{Thuan \& Izotov(2005)}]{Thuan05}
Thuan, T.~X. \& Izotov, Y.~I. 2005, \apjs, 161, 240

\bibitem[{Thuan {et~al.}(1997)Thuan, Izotov, \& Lipovetsky}]{Thuan97}
Thuan, T.~X., Izotov, Y.~I., \& Lipovetsky, V.~A. 1997, \apj, 477, 661

\bibitem[{Trebitsch {et~al.}(2017)Trebitsch, Blaizot, Rosdahl, Devriendt, \&
  Slyz}]{Trebitsch17}
Trebitsch, M., Blaizot, J., Rosdahl, J., Devriendt, J., \& Slyz, A. 2017,
  \mnras, 470, 224

\bibitem[{Veilleux \& Osterbrock(1987)}]{Veilleux87}
Veilleux, S. \& Osterbrock, D.~E. 1987, \apjs, 63, 295

\bibitem[{Weaver {et~al.}(1977)Weaver, McCray, Castor, Shapiro, \&
  Moore}]{Weaver77}
Weaver, R., McCray, R., Castor, J., Shapiro, P., \& Moore, R. 1977, \apj, 218,
  377

\bibitem[{Weilbacher {et~al.}(2012)Weilbacher, Streicher, Urrutia, Jarno,
  P{\'e}contal-Rousset, Bacon, \& B{\"o}hm}]{Weilbacher12}
Weilbacher, P.~M., Streicher, O., Urrutia, T., {et~al.} 2012, in SPIE
  Astronomical Telescopes + Instrumentation, ed. N.~M. Radziwill \& G.~Chiozzi
  (SPIE), 84510B

\bibitem[{Westmoquette {et~al.}(2009)Westmoquette, Gallagher, Smith, Trancho,
  Bastian, \& Konstantopoulos}]{Westmoquette09}
Westmoquette, M.~S., Gallagher, J.~S., Smith, L.~J., {et~al.} 2009, \apj, 706,
  1571

\bibitem[{Williamson {et~al.}(2016)Williamson, Martel, \&
  Kawata}]{Williamson16}
Williamson, D., Martel, H., \& Kawata, D. 2016, \apj, 822, 91

\bibitem[{Zackrisson {et~al.}(2011)Zackrisson, Rydberg, Schaerer, {\"O}stlin,
  \& Tuli}]{Zackrisson11}
Zackrisson, E., Rydberg, C.-E., Schaerer, D., {\"O}stlin, G., \& Tuli, M. 2011,
  \apj, 740, 13

\bibitem[{Zastrow {et~al.}(2013)Zastrow, Oey, Veilleux, \&
  McDonald}]{Zastrow13}
Zastrow, J., Oey, M.~S., Veilleux, S., \& McDonald, M. 2013, \apj, 779, 76

\bibitem[{Zastrow {et~al.}(2011)Zastrow, Oey, Veilleux, McDonald, \&
  Martin}]{Zastrow11}
Zastrow, J., Oey, M.~S., Veilleux, S., McDonald, M., \& Martin, C.~L. 2011,
  \apj, 741, L17

\end{thebibliography}

\end{document}